\documentclass[12pt,preprint]{aastex}

\begin{document}
 \title{Hierarchical star formation in the Milky Way disk} 

 \author{R.~de~la~Fuente~Marcos and C. de la Fuente Marcos}
  \affil{Suffolk University Madrid Campus, C/ Vi\~na 3,
         E-28003, Madrid, Spain}
  \email{raul@galaxy.suffolk.es}

  \begin{abstract}
     Hierarchical star formation leads to a progressive decrease in 
     the clustering of star clusters both in terms of spatial scale 
     and age. Consistently, the statistical analysis of positions 
     and ages of clusters in the Milky Way disk strongly suggests 
     that a correlation between the duration of star formation in a 
     region and its size does exist. The average age difference 
     between pairs of open clusters increases with their separation 
     as the $\sim$0.16 power. In contrast and for the Large Magellanic 
     Cloud, Efremov \& Elmegreen (1998) found that the age difference 
     scales with the $\sim$0.35 power of the region size. This 
     discrepancy may be tentatively interpreted as an argument in 
     support of intrinsically shorter (faster) star formation 
     time-scales in smaller galaxies. However, if both the effects of 
     cluster dissolution and incompleteness are taken into 
     consideration, the average age difference between cluster pairs 
     in the Galaxy increases with their separation as the $\sim$0.4 
     power. This result implies that the characteristic time-scale 
     for coherent, clustered-mode star formation is nearly 1 Myr. 
     Therefore, the overall consequence of ignoring the effect of 
     cluster dissolution is to overestimate the star formation 
     time-scale. On the other hand, in the Galactic disk and for young 
     clusters separated by less than three times the characteristic 
     cluster tidal radius (10 pc), the average age difference is 16 
     Myr, which suggests common origin. A close pair classification 
     scheme is introduced and a list of 11 binary cluster candidates 
     with physical separation less than 30 pc is compiled. Two of 
     these pairs are likely primordial: ASCC 18/ASCC 21 and NGC 3293/NGC 3324. A 
     triple cluster candidate in a highly hierarchical configuration 
     is also identified: NGC 1981/NGC 1976/Collinder 70 in Orion. We 
     find that binary cluster candidates seem to show a tendency to 
     have components of different size; an evidence for dynamical 
     interaction.
  \end{abstract}


  \section{Introduction}
    Star clusters do not form in isolation but tend to be clustered themselves 
    in complexes (Efremov 1978). Efremov \& Elmegreen (1998) studied the
    positions and ages of Cepheid variables and star clusters in the Large
    Magellanic Cloud (LMC) and found that objects closer to each other had 
    also a substantially higher probability of being almost coeval. This
    finding has been interpreted as evidence in favor of star formation 
    proceeding faster in smaller regions than in larger ones. This idea was 
    originally proposed by Elmegreen et al. (1996). In their work and using a 
    sample of Magellanic spiral and irregular galaxies and blue compact dwarfs,
    they concluded that star complexes are systematically smaller (in absolute
    terms) in smaller galaxies. An orderly decrease in the star formation 
    time-scale with galaxy size was found. From a theoretical perspective, this 
    is to be expected as gravitational collapse is started when the local 
    free-fall time becomes shorter than the sound-crossing time with the latter
    being directly proportional to the size of the region undergoing collapse.
    The star formation rate essentially equals the self-gravity rate and the
    timing for star formation is hierarchical, with a number of small active 
    regions being born and dissolved in the time it takes the larger region
    surrounding them to finish (Elmegreen 2000). Star cluster complexes are
    the largest groups in the hierarchy of star formation. In quantitative terms,  
    Efremov \& Elmegreen (1998) concluded that the average age difference
    between pairs of star clusters in the LMC increased with their separation 
    as the $\sim$0.35 power, following the law:
    \begin{equation}
       \Delta t ({\rm Myr}) \sim 3.3 \ S({\rm pc})^{0.35}\,,
        \label{LMC}
    \end{equation} 
    for clusters in the 0.01-1 deg separation range (8-780 pc at the assumed
    LMC distance, 45 kpc). The catalogue of LMC clusters used in their research
    was compiled by Bica et al. (1996). In summary, star formation in the 
    LMC is hierarchical in space and time (Efremov \& Elmegreen 1998). If the 
    interpretation envisaged by Efremov \& Elmegreen (1998) is correct, we 
    would expect slower star formation and, therefore, a smaller value for the 
    scaling power index in the Milky Way disk. In this research we attempt to 
    confirm this expectation. As in Efremov \& Elmegreen (1998), the results 
    obtained here are only applicable to the so-called clustered mode of star 
    formation.

    This paper is organized as follows: in \S 2, we present the open cluster 
    sample used in this research and study the cluster age difference versus
    separation. The closest pairs are presented in \S 3. In \S 4 we discuss 
    our results and finally, in \S 5, we summarize our conclusions.  

  \section{Open cluster age difference versus separation} 
    In this paper, a sample of open clusters has been extracted from the 
    {\it Open Cluster Database} \footnote{http://www.univie.ac.at/webda/} 
    (WEBDA, Mermilliod \& Paunzen 2003). The latest update of WEBDA (April 2009, 
    Paunzen \& Mermilliod 2009) includes 1028 open clusters with both age and 
    distance known (out of a total of 1756 objects). This is the sample used 
    to study the hierarchical properties of star formation in the Milky Way 
    disk. In principle, WEBDA does not contain stellar associations or 
    embedded clusters. Following Efremov \& Elmegreen (1998), we consider all 
    cluster pairs from the sample under study (527,878) and obtain their 
    physical separation (in the usual metric) in pc and age difference in Myr. 
    Then and within certain age ranges, we calculate the average age difference 
    $\Delta t$ among these clusters as a function of their physical separation 
    $S$ for regular intervals of separation (50 pc). As for the age difference, 
    the absolute value is computed. In Efremov \& Elmegreen (1998) the 
    intercluster separation is defined to be the de-projected distance between 
    the two clusters, considering zero depth to the LMC and an inclination of 
    33$^{\circ}$. This is a major difference with respect to the definition 
    used in our present work: here we use full spatial separations. Open 
    clusters form in star complexes. Efremov (1978) gave the first quantitative
    definition of the term Star Complex, {\it vast aggregates of stars with an 
    average diameter of $\approx$ 600 pc and an age of tens of millions of 
    years, encompassing stars that originated in the same gas-dust complex}. 
    Elmegreen (2009) has shown that the clump scale for giant cloud formation
    is $\sim$600 pc for local galaxies; at larger redshifts, however, the clump 
    scale is larger, $\sim$1500 pc. This is why, in our analysis, we focus on the 
    pair subsample with separations $\leq$ 600 pc (31641 pairs). As the probability 
    of finding a cluster pair (of random ages) with separation $\leq$ 600 pc is 
    0.06, it may be argued that the 600 pc size-scale is arbitrary as most open 
    clusters do not form within complexes. This concern can be, however, neglected. 
    The characteristic time-scale for stars to become part of the field stellar 
    populations is 10-20 Myr (e.g. Battinelli \& Capuzzo-Dolcetta 1989, 1991; Lada
    \& Lada 1991, 2003; Kroupa \& Boily 2002; de la Fuente Marcos \& de la Fuente 
    Marcos 2004, 2008; Bastian et al. 2005; Fall et al. 2005). If we restrict our 
    analysis to open clusters younger than 20 Myr (212 clusters), we observe that 
    67\% of them have at least one neighbor in the same age group within 300 pc, 
    86\% of them within 600 pc, and 93\% within 1000 pc. Therefore, the present-day
    fraction of open clusters formed away from complexes in the Galactic disk appears 
    to be rather negligible.

\begin{figure}
        \epsscale{0.49}
        \plotone{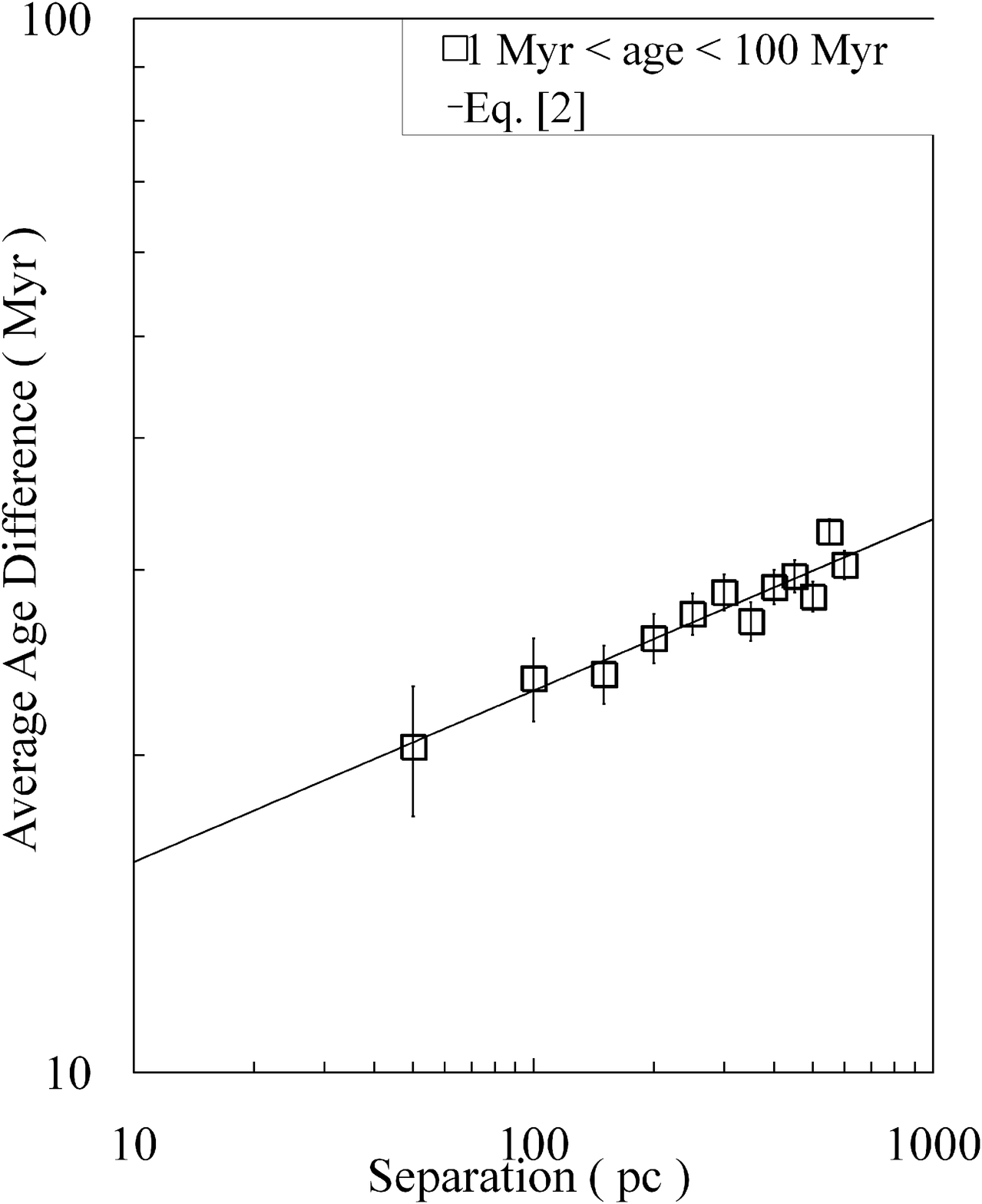}
        \plotone{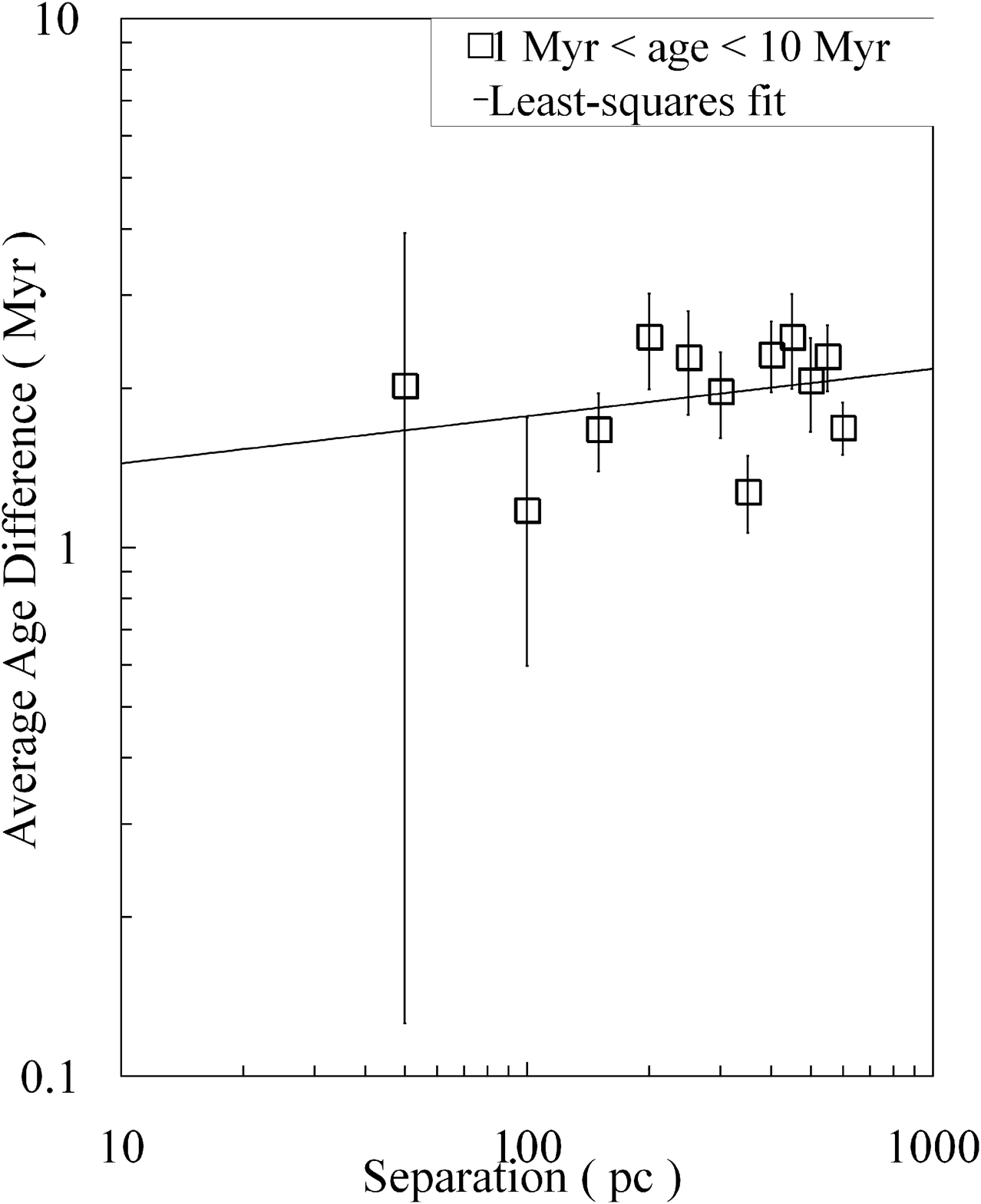}
        \plotone{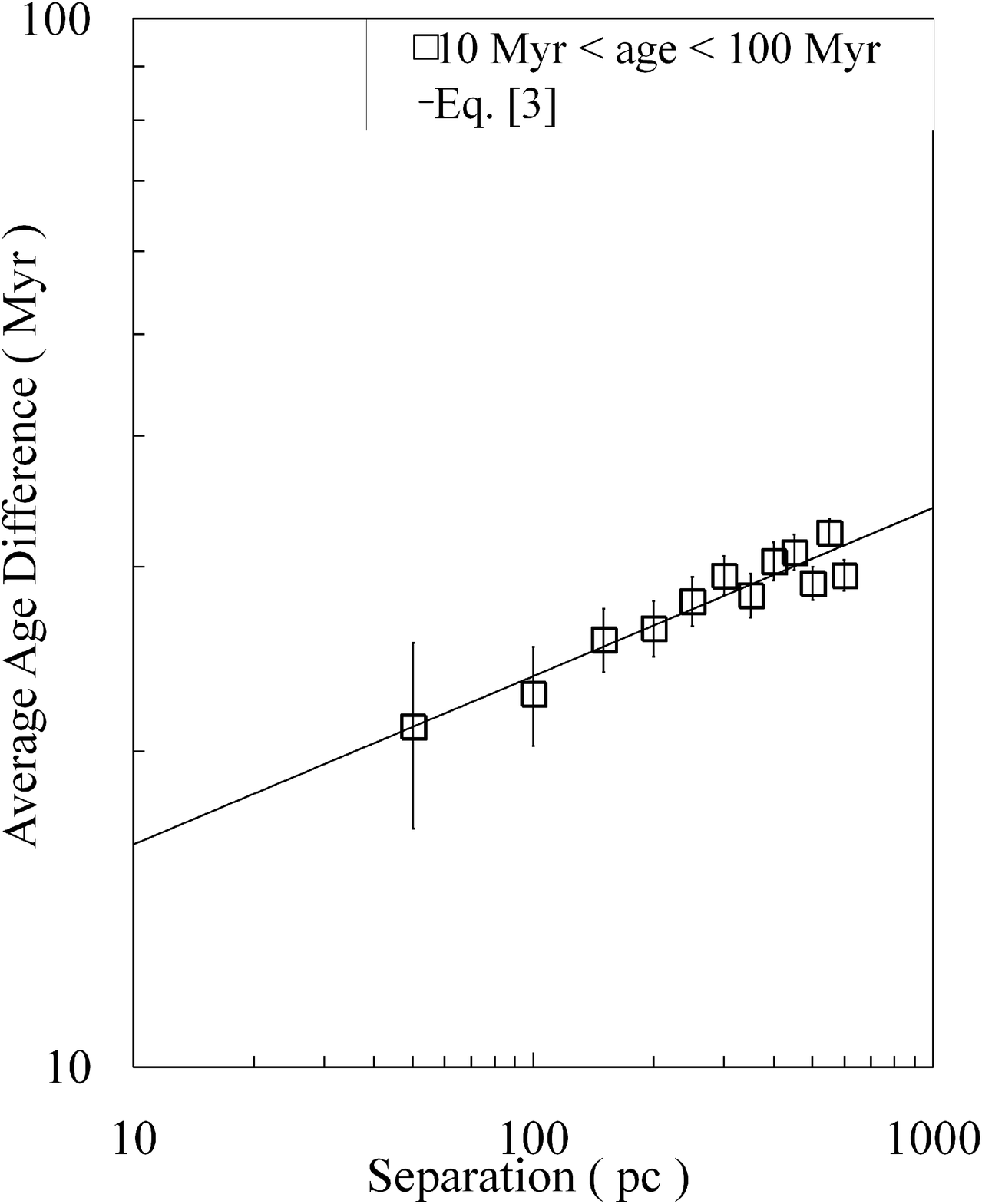}
        \plotone{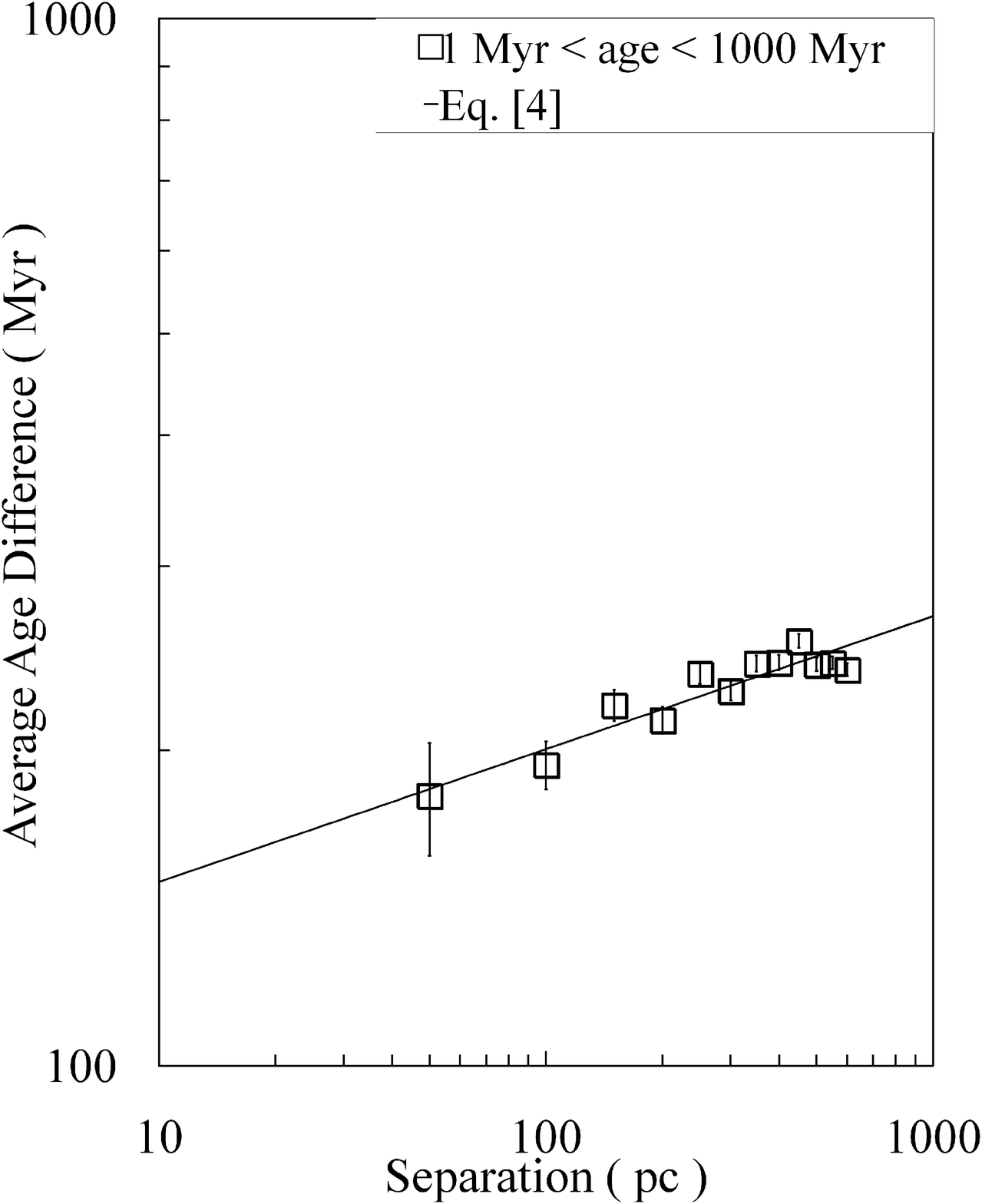}
        \caption{Average age difference between open cluster pairs as a
                 function of their physical separation. Following Efremov
                 \& Elmegreen (1998), four age intervals have been 
                 considered. These figures clearly show that the age difference 
                 is smaller for closer open clusters. Error bars display the
                 standard error in the mean.
                 \label{aad}
                }
     \end{figure}

    The results of the calculation described above are shown in Figure \ref{aad}. 
    Error bars display the standard error in the mean. As in Efremov \& Elmegreen 
    (1998) and in order to facilitate direct comparison, four age intervals are 
    considered: 1 to 100 Myr, 1 to 10 Myr, 10 to 100 Myr, and 1 to 1000 Myr. The 
    average age difference between pairs of open clusters increases systematically 
    with their spatial separation. The number of clusters pairs within these age 
    intervals is 5286, 241, 3637, and 27142, respectively. The least-squares 
    fits in the $S$ = 0-600 pc separation range for the, statistically 
    significant, $\Delta t-S$ relations shown in Figure \ref{aad} are:
    \begin{equation}
       \log \Delta t ({\rm Myr}) = 1.04\pm0.04 \ + 
                                   0.163\pm0.015 \ \log S({\rm pc}), \ \ \
               {\rm 1-100 \ Myr}\,,
    \end{equation}
    \begin{equation}
       \log \Delta t ({\rm Myr}) = 1.05\pm0.04 \ + 
                                   0.160\pm0.016 \ \log S({\rm pc}), \ \ \
               {\rm 10-100 \ Myr}\,,
    \end{equation}
    \begin{equation}
       \log \Delta t ({\rm Myr}) = 2.05\pm0.04 \ + 
                                   0.127\pm0.015 \ \log S({\rm pc}), \ \ \
               {\rm 1-1000 \ Myr}\,.
    \end{equation}
    The correlation coefficients for these three fits are 0.96, 0.96, and 0.94, 
    respectively. The 1-10 Myr data interval results are not statistically
    significant (low number of open cluster pairs and correlation coefficient
    of 0.27). On the other hand, cluster ages $\leq$ 10 Myr are highly unreliable.
    The probable uncertainties of the parameters of the straight-line
    fit have been calculated as described in Press et al. (2002), page 664.
    The goodness of fit has been estimated using the Pearson product-moment 
    correlation coefficient ($r$ = 1 for perfect fit, e.g. Wall \& Jenkins 
    2003). The determination coefficient ($r^2$) yields the proportion of the
    variance in the dependent variable that can be explained by the regression
    equation.

\begin{figure}
        \epsscale{0.49}
        \plotone{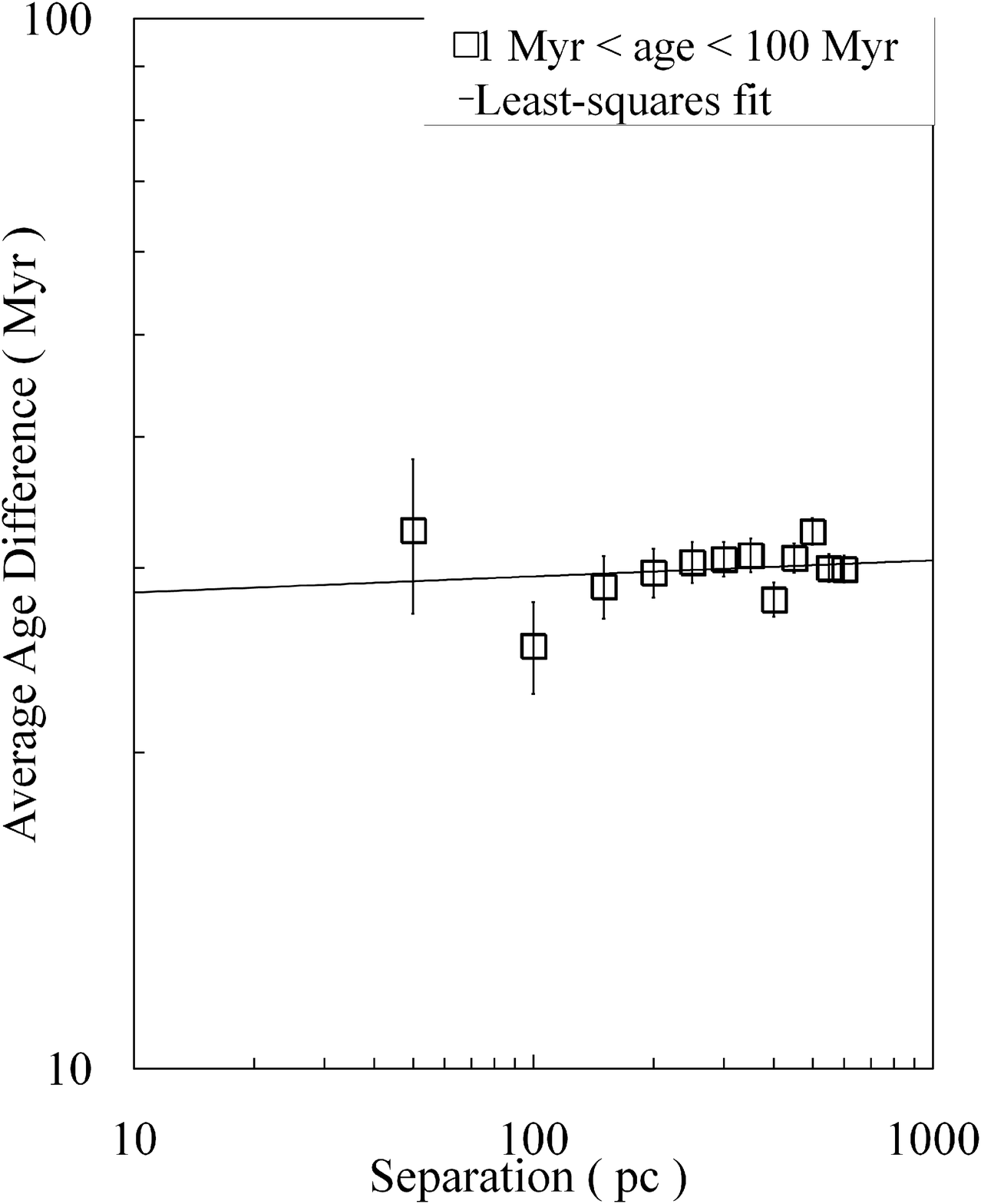}
        \plotone{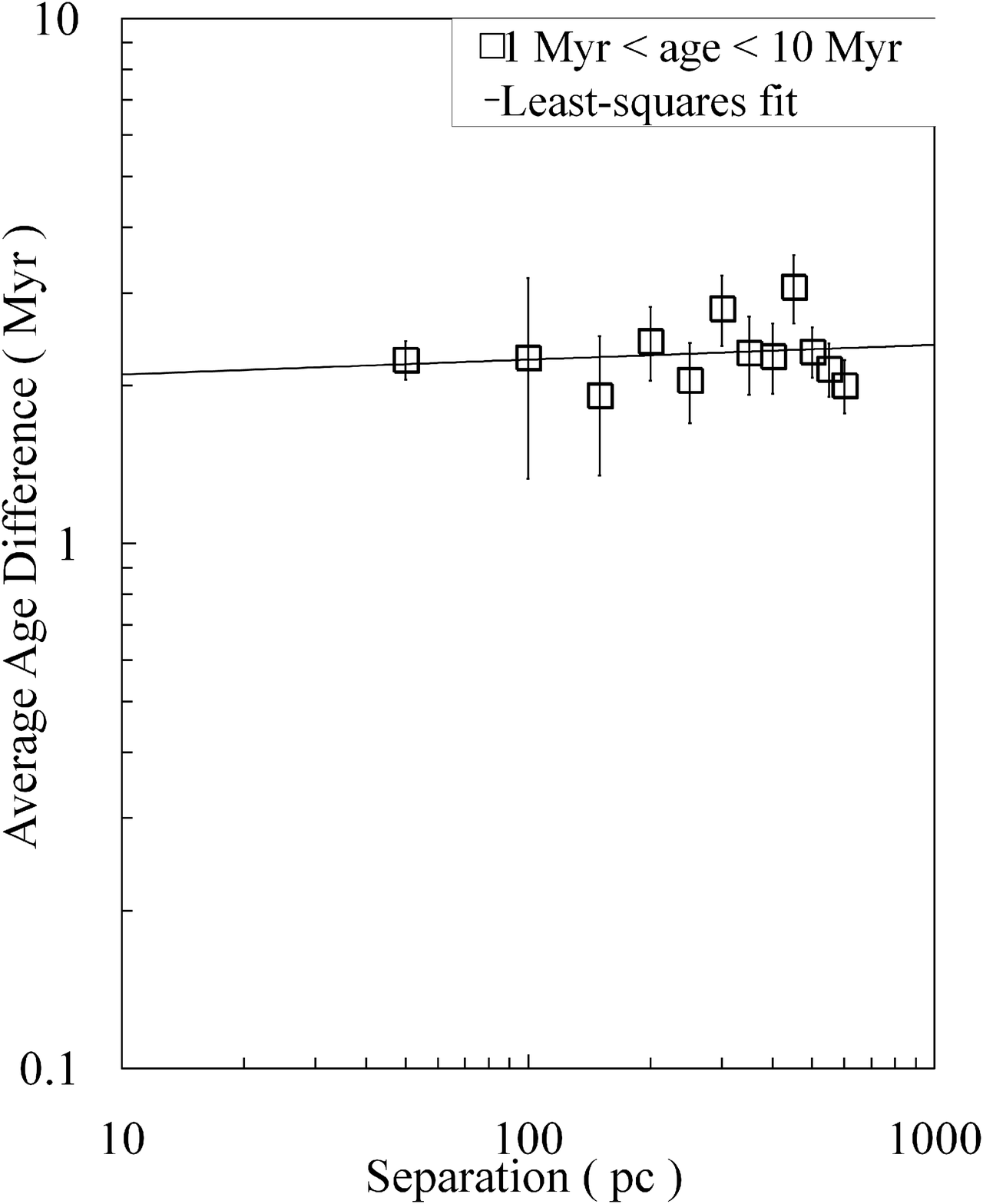}
        \plotone{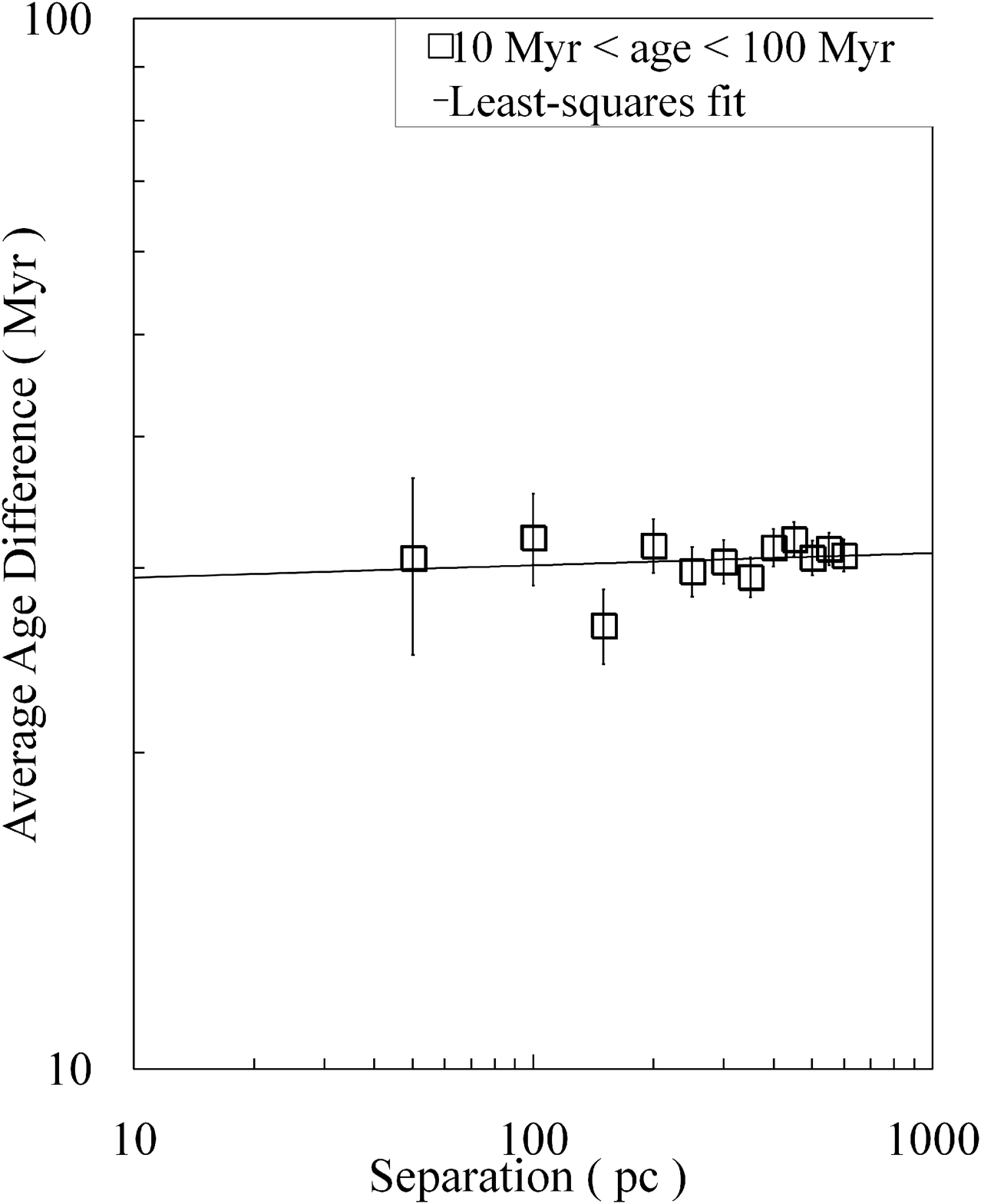}
        \plotone{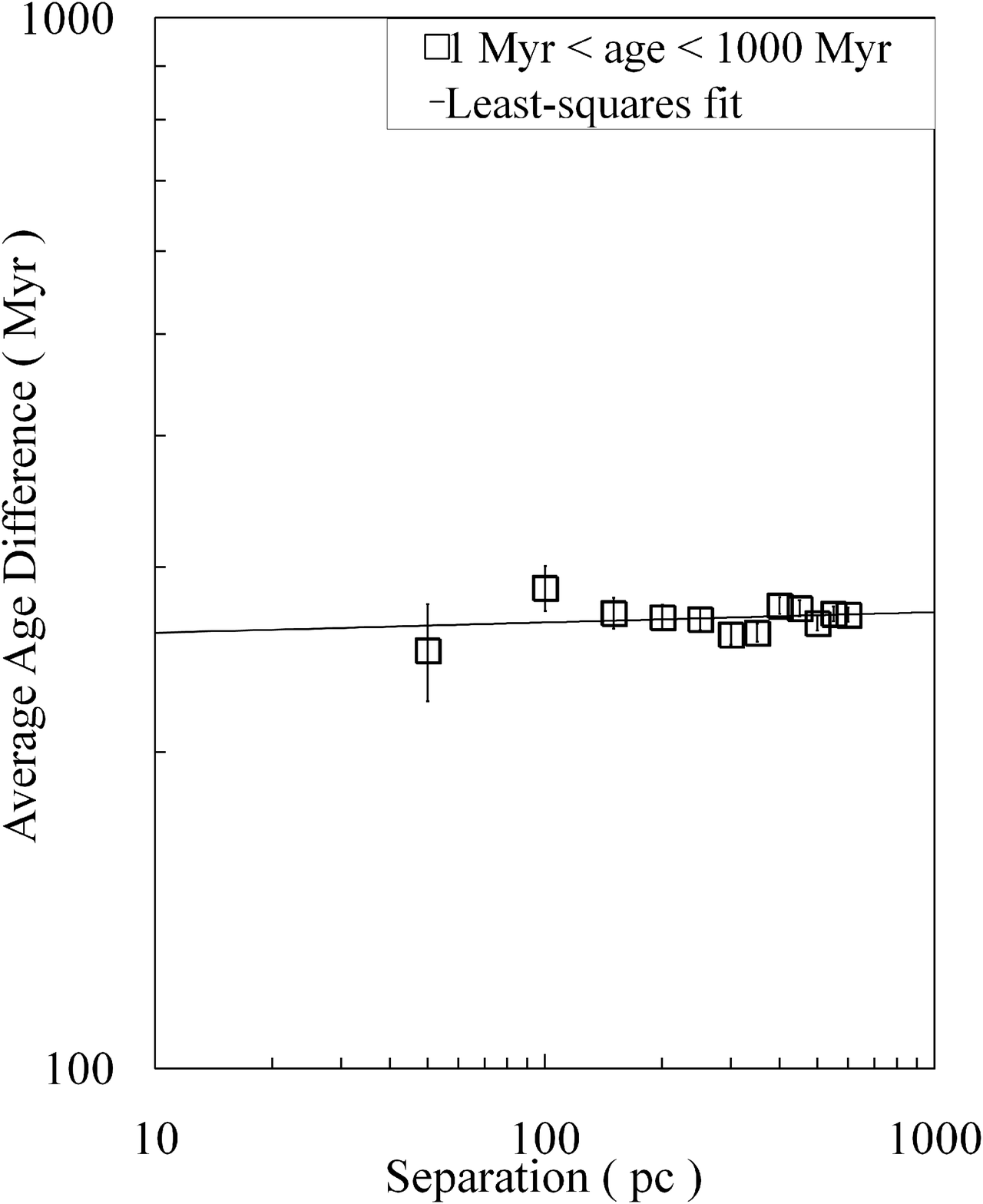}
        \caption{Same as Figure \ref{aad} but with open cluster ages scrambled 
                 randomly among all the 1028 objects in the sample. The
                 correlation disappears for randomized ages, indicating that 
                 the correlation found for the actual data is statistically
                 robust. This is just one example out of 50,000 trials (see
                 Figure \ref{trials}) with randomly scrambled data and real 
                 cluster positions.
                 \label{exp}
                }
     \end{figure}

\begin{figure}
        \epsscale{0.49}
        \plotone{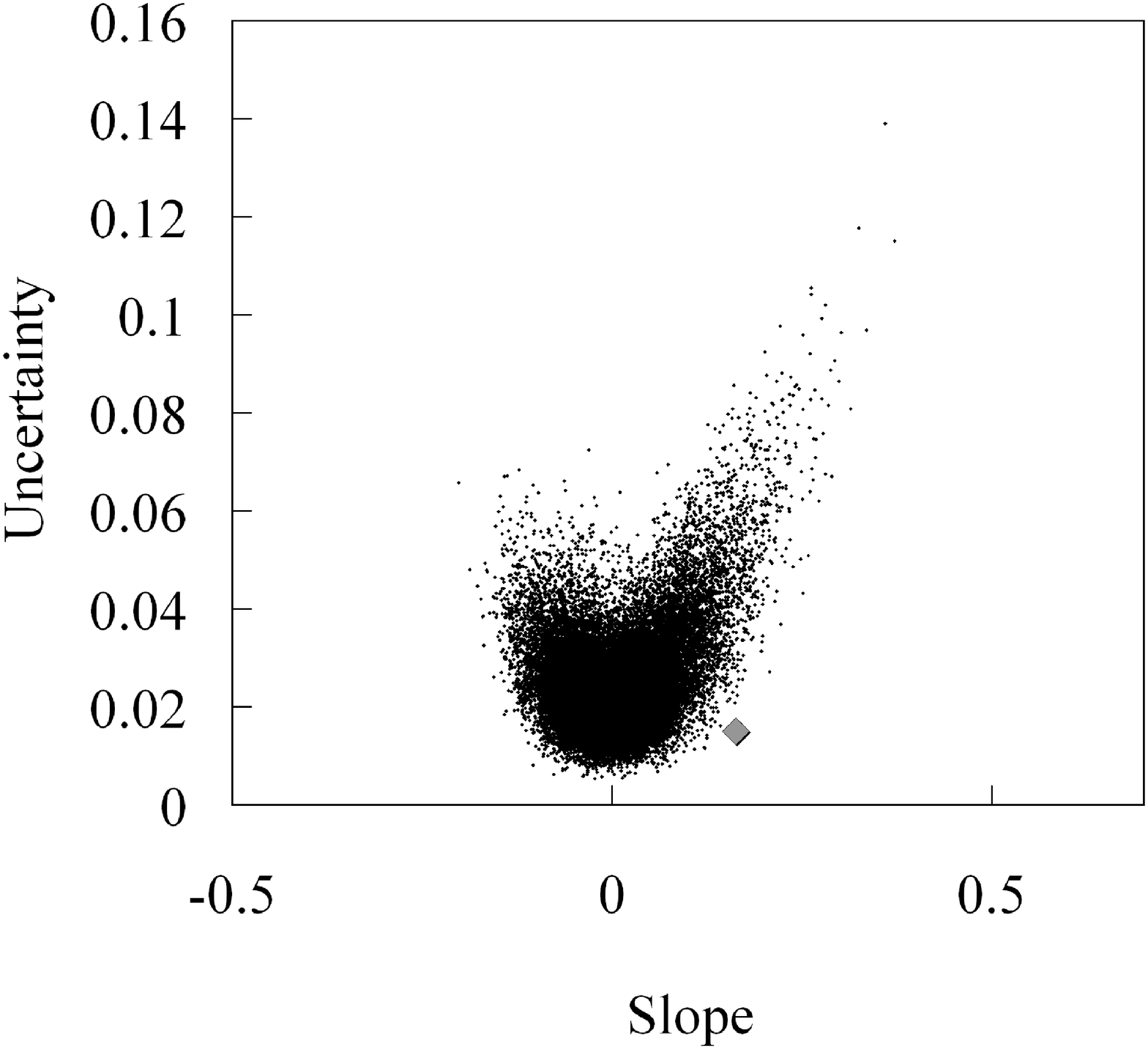}
        \plotone{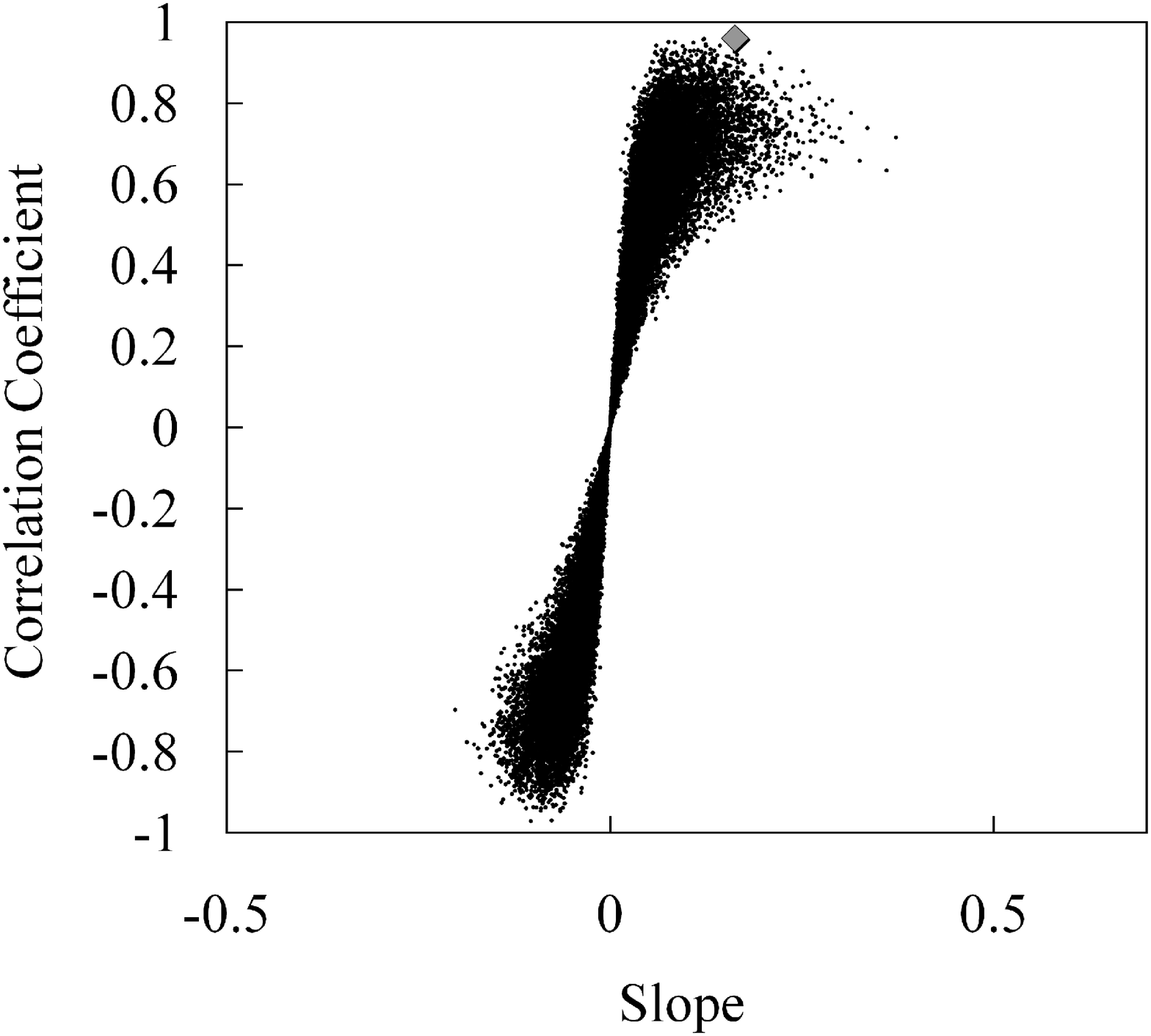}
        \plotone{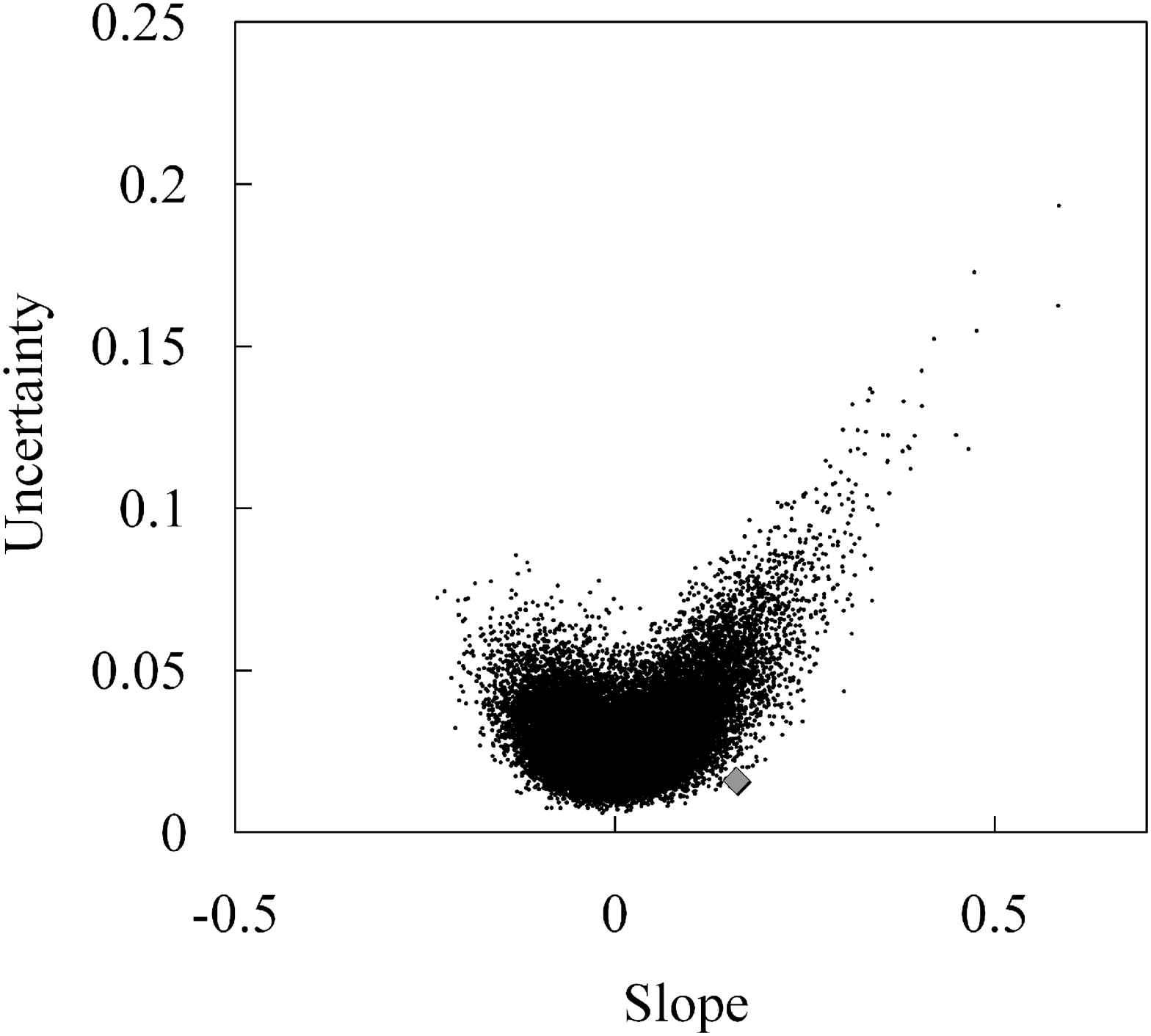}
        \plotone{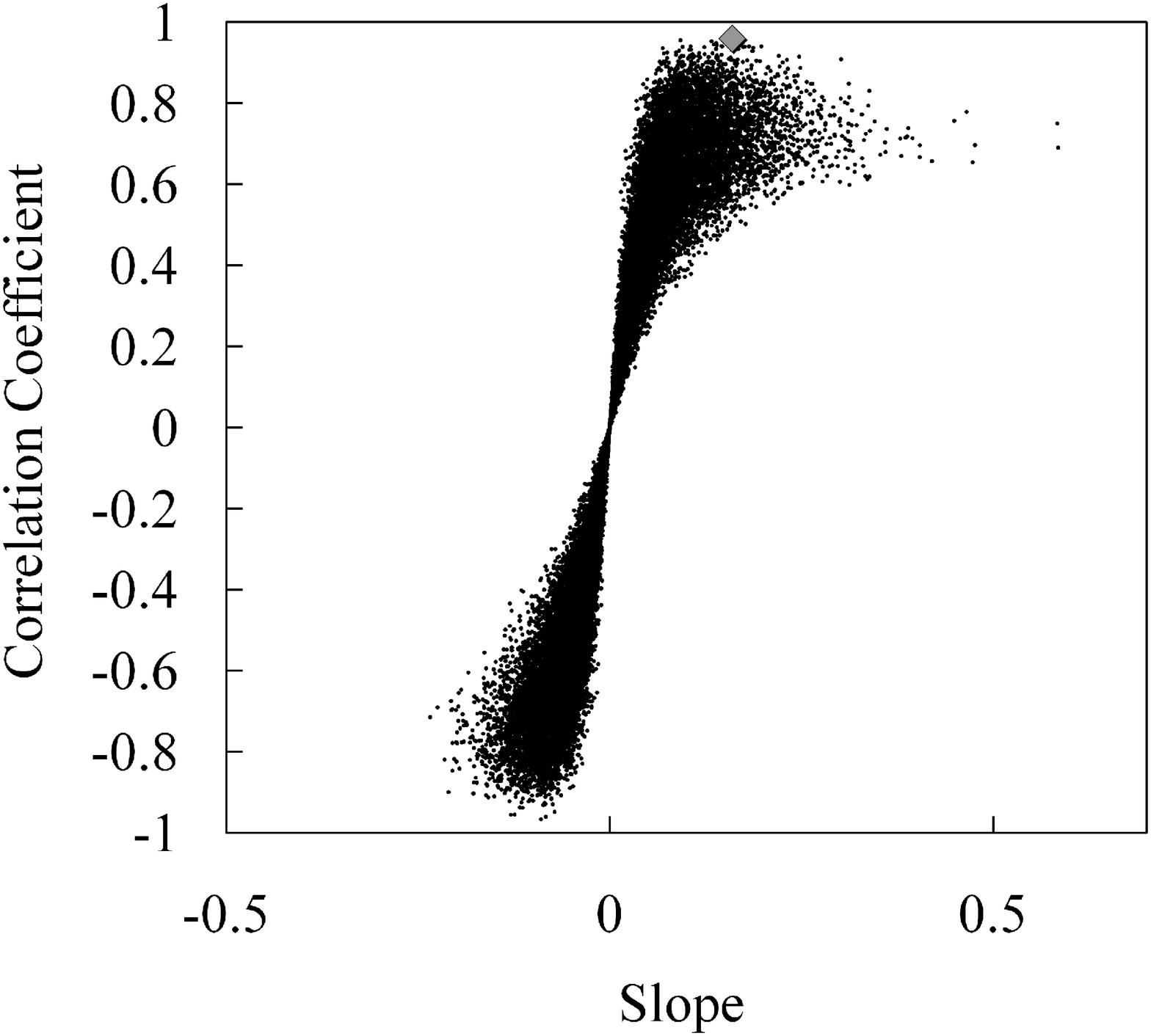}
        \caption{Could the apparent correlation in Figure \ref{aad} be a
                 statistical artifact? In order to confirm that the 
                 correlation found is statistically significant, we have 
                 completed 50,000 trials with randomly scrambled age data 
                 and real open cluster positions. In the top panels we 
                 represent the uncertainty in the slope as a function of 
                 the slope (left) for both the simulated trials and the 
                 actual result for the 1-100 Myr age range. Also displayed 
                 is the correlation coefficient as a function of the slope
                 (right). The bottom panels show the same data for the
                 10-100 Myr age range. The original results are displayed
                 as black diamond symbols. The probability of finding a
                 random trial with slope within 1 $\sigma$ of the value
                 obtained for the unaltered data set and with correlation 
                 coefficient $>$ 0.95 is 0.00002 (1 favorable case in 
                 50,000 trials); the original correlation coefficient is 
                 0.96.  This result clearly indicates that the open cluster
                 $\Delta t-S$ correlation is statistically robust.
                 \label{trials}
                }
     \end{figure}

    But, could the apparent correlation obtained from the above equations be a
    statistical artifact? What is the likelihood of getting by chance a joint
    combination of average age difference and cluster pair physical separation that
    would be correlated 0.96 or greater? A statistical test of randomly scrambled 
    age data and real open cluster positions similar to the one used in Efremov \& 
    Elmegreen (1998) confirms that the correlation found above is statistically
    robust as it disappears for randomized data. We have completed 50,000
    trials with randomly scrambled age data and real open cluster positions; 
    the real cluster ages are reassigned randomly to different clusters. Figure 
    \ref{exp} is analogous to Figure \ref{aad} and shows the results of one
    of these trials. The correlations disappear as the average age difference
    tends to be the same regardless of separation. Figure \ref{trials} provides 
    the results of the entire set of 50,000 trials. Two plots are displayed: the 
    uncertainty in the slope as a function of the slope itself (left) and the 
    correlation coefficient as a function of the slope (right). The top panels 
    correspond to the 1-100 Myr age range and the bottom panels show the same 
    results for the 10-100 Myr age range. The original results are also displayed
    as black diamond symbols. For the 1-100 Myr age range, the probability of 
    finding a random trial with slope within 1 $\sigma$ of the value obtained for 
    the real data is 0.010. If we also impose that the correlation coefficient is 
    $>$ 0.8, that probability decreases to 0.0026. If the correlation coefficient 
    has to be $>$ 0.9 we find 15 favorable cases. For $r>$0.95 only 1 favorable case 
    is found out of 50,000 trials. The probability of getting by chance such a 
    correlation (ours was 0.96), or higher, is completely negligible. This result 
    clearly indicates that the open cluster $\Delta t-S$ correlation is statistically 
    robust. 

    On the other hand, the fact that Figure \ref{trials} exhibits a very noticeable 
    asymmetry may hint of a data processing problem or data quality issue and argue 
    against the validness of the analysis presented in the previous paragraph. Even 
    with the randomized age trials, a positive slope is favored in the displayed
    age ranges. There is however an obvious reason for this asymmetry: for an open 
    cluster sample in the range 1-100 Myr, the number of objects in the age range 1-20 Myr 
    is nearly 50\% of the sample size (see, e.g., Fig. 2 in de la Fuente Marcos \&
    de la Fuente Marcos 2008). For a young cluster, the probability of getting a 
    reassigned age that is relatively close to its actual age is rather high. The
    rapid decrease in cluster numbers for ages older than 20 Myr ({\it infant 
    mortality}) is the result of catastrophic gas ejection (e.g. Boily \& Kroupa 2003;
    Goodwin \& Bastian 2006), a process that, although frequently ignored, was first 
    proposed by Hills (1980). The asymmetry is greatly reduced if the age range
    1-1000 Myr is considered. The observed asymmetry can be understood as the signature 
    of early cluster disruption. The effect is also apparent in Efremov \& Elmegreen 
    (1998, Fig. 3) but no explanation is given there.

    The sample considered above is, in principle, not volume-limited and it may be 
    argued that it is strongly biased in favor of young objects as older open clusters 
    are less likely to be included because they are more difficult to identify. 
    Completeness of general open cluster samples has been traditionally approached under 
    the assumption of uniform surface density of open clusters in the Solar Neighborhood 
    (Battinelli \& Capuzzo-Dolcetta 1989, 1991). This hypothesis implies $N \propto d^2$, 
    where $N$ is the number of clusters and $d$, a given heliocentric distance. In their 
    papers, it was found that for open clusters within 2 kpc from the Sun and brighter than
    $M_V$ = -4.5, the assumption of uniform average number density of open clusters
    was matched well by the observational results. Their sample of 100 objects included
    clusters younger than 1.6 Gyr (Battinelli et al. 1994). More recently, Piskunov et 
    al. (2006) have concluded that assuming uniform density the completeness limit for 
    clusters of any age could be 0.85 kpc. We found similar results in our sample: for 
    objects in the age range 1-1000 Myr located within 0.9 pc from the Sun the power-law 
    index is 1.96$\pm$0.05 with a correlation coefficient of 0.998. For clusters in the 
    age range 200-1000 Myr located within 1 kpc from the Sun with index 1.96$\pm$0.06 
    and $r$ = 0.996. Our sample is likely to be at least 90\% complete for older clusters 
    if the radial distance is restricted to 1 kpc. For open clusters with $d <$ 2 kpc 
    \begin{equation}
       \log \Delta t ({\rm Myr}) = 1.02\pm0.04 \ + 
                                   0.176\pm0.016 \ \log S({\rm pc}), \ \ \
               {\rm 1-100 \ Myr}\,,
    \end{equation}
    and for $d <$ 1 kpc
    \begin{equation}
       \log \Delta t ({\rm Myr}) = 1.07\pm0.07 \ + 
                                   0.16\pm0.03 \ \log S({\rm pc}), \ \ \
               {\rm 1-100 \ Myr}\,.
    \end{equation}
    Not surprisingly, our volume-limited results are fully consistent, within the error 
    limits, with those from the general sample discussed above as $S$ = 0-600 pc in
    all the calculations. The correlation coefficients are 0.96 and 0.87, respectively. 
    Therefore, our initial sample was already, from a certain point of view, volume-limited 
    and selection effects were rather negligible.

    The correlation coefficients for the sample studied here are quite good and
    the results of the random trials are consistent, indicating that the open 
    cluster $\Delta t$-$S$ correlation is statistically significant. Following 
    Efremov \& Elmegreen (1998), we can conclude that open clusters in the 
    Milky Way disk form in a hierarchical sequence in which the duration of 
    star formation in a region scales with the $\sim0.16\pm0.02$ power of the 
    region size over scales smaller than 600 pc. The error quoted here is the result
    of a formal least-squares fit and not the absolute uncertainty. The correlation 
    can be written as:
    \begin{equation}
       \Delta t ({\rm Myr}) \sim 11.1 \ S({\rm pc})^{0.16}\,.
        \label{rel}
    \end{equation} 
    The difference found between the Milky Way and the LMC (Equation \ref{LMC} 
    vs. Equation \ref{rel}) regarding the slope of the correlation is therefore
    statistically significant. Could it be the result of different galaxy type?
    After all, the Milky Way is a disk galaxy dominated by a spiral arm 
    structure. The LMC is classified as a barred Magellanic spiral galaxy 
    (SBms) and is morphologically in between normal spiral galaxies and irregular 
    galaxies. It has one spiral arm and a bar. The LMC displays a very prominent bar 
    near its center, suggesting that it may have previously been a barred spiral 
    galaxy. A significant number of candidate binary clusters in the LMC appear to 
    be associated with the bar (e.g. Leon et al. 1999). Most clusters appear to be 
    associated with the bar and a surrounding structure that could be the result of 
    the interaction between the LMC and the SMC (Bica et al. 1996). The LMC bar is a 
    region with increased cluster density and 46\% of all bar clusters can be found 
    in close pairs or multiple systems (Dieball \& Grebel 1999; Dieball et al. 2002). 
    The survival time for cluster pairs is longer in the bar region where the tidal
    field of the LMC is probably weaker (Bhatia 1990; Elson et al. 1987).
    Schommer et al. (1992) found that the LMC cluster system has kinematics consistent 
    with the clusters moving in a disk-like distribution. Grocholski et al. (2007) 
    have confirmed the existence of a single rotating disk with clusters and field 
    stars lying in the same plane. The outer clusters also exhibit disk kinematics and 
    there are clear signs of perturbations from the bar. The dominant effect of structure 
    evolution in the LMC is general galactic dynamics (Bastian et al. 2009). With this 
    scenario in mind, the formation and early dynamics of young clusters in the LMC may 
    be indeed similar to what is observed in the inner disk of our own Galaxy, effect of 
    the bar included. There is clear evidence of faster star formation in the direction 
    of Scutum and Norma (de la Fuente Marcos \& de la Fuente Marcos 2008) in the Milky Way 
    disk. The results provided above seem to suggest that the different LMC and Milky Way star
    formation morphologies have a minor role in the overall organization of the spatial 
    correlation of star formation. Certainly, the different strength of the spiral patterns 
    may have some effect but it appears not to be dominant. 

\begin{figure}
        \epsscale{0.32}
        \plotone{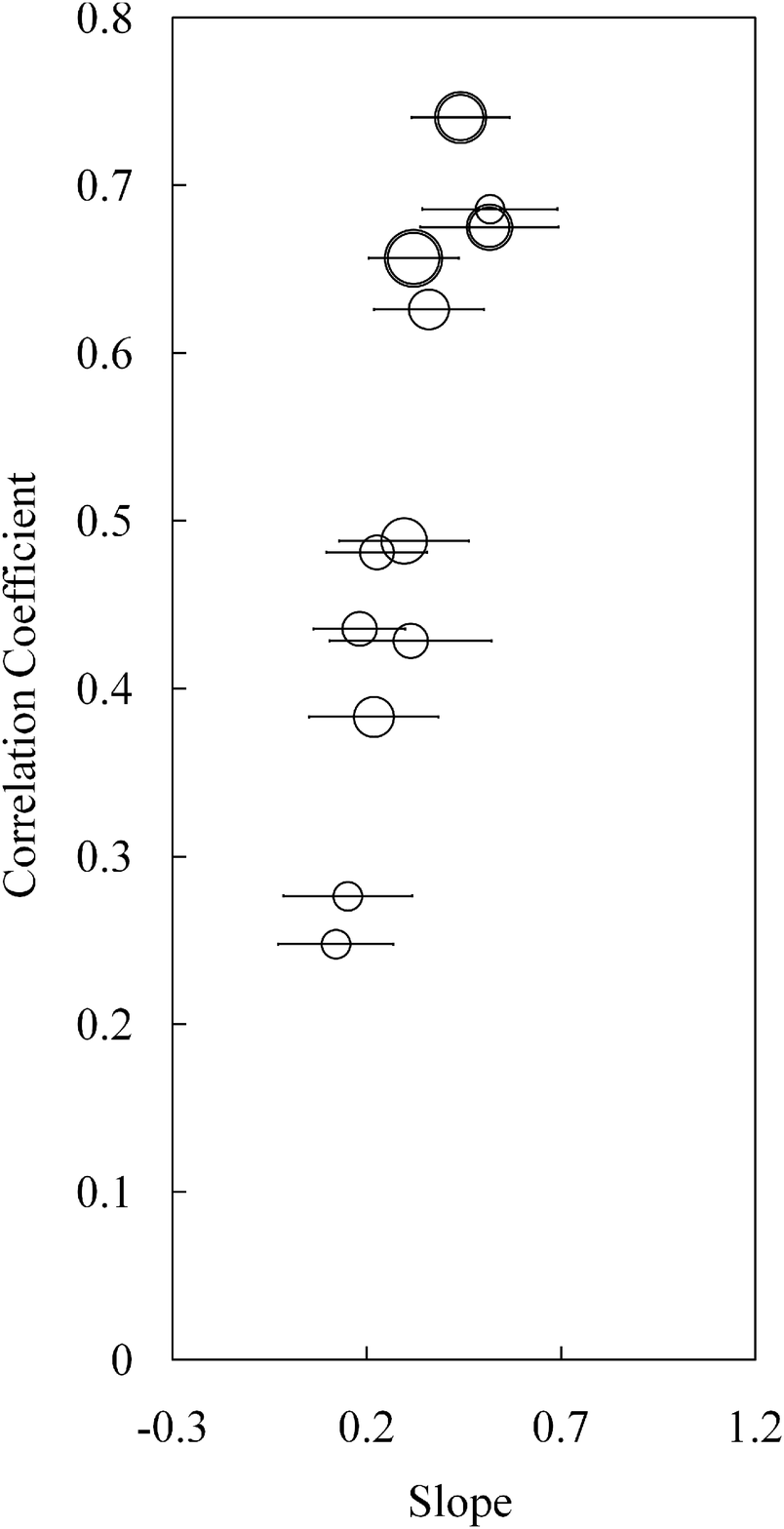}
        \plotone{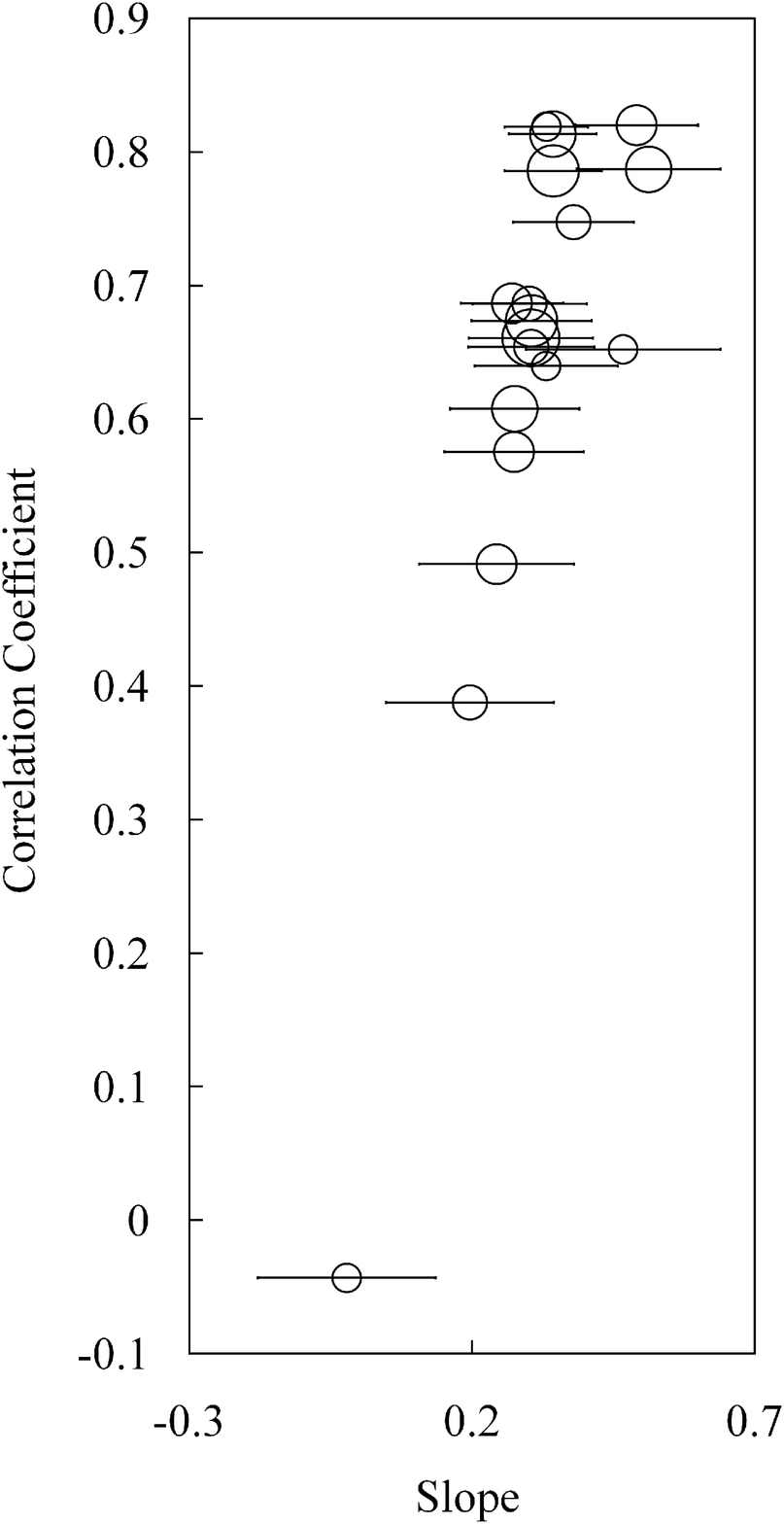}
        \plotone{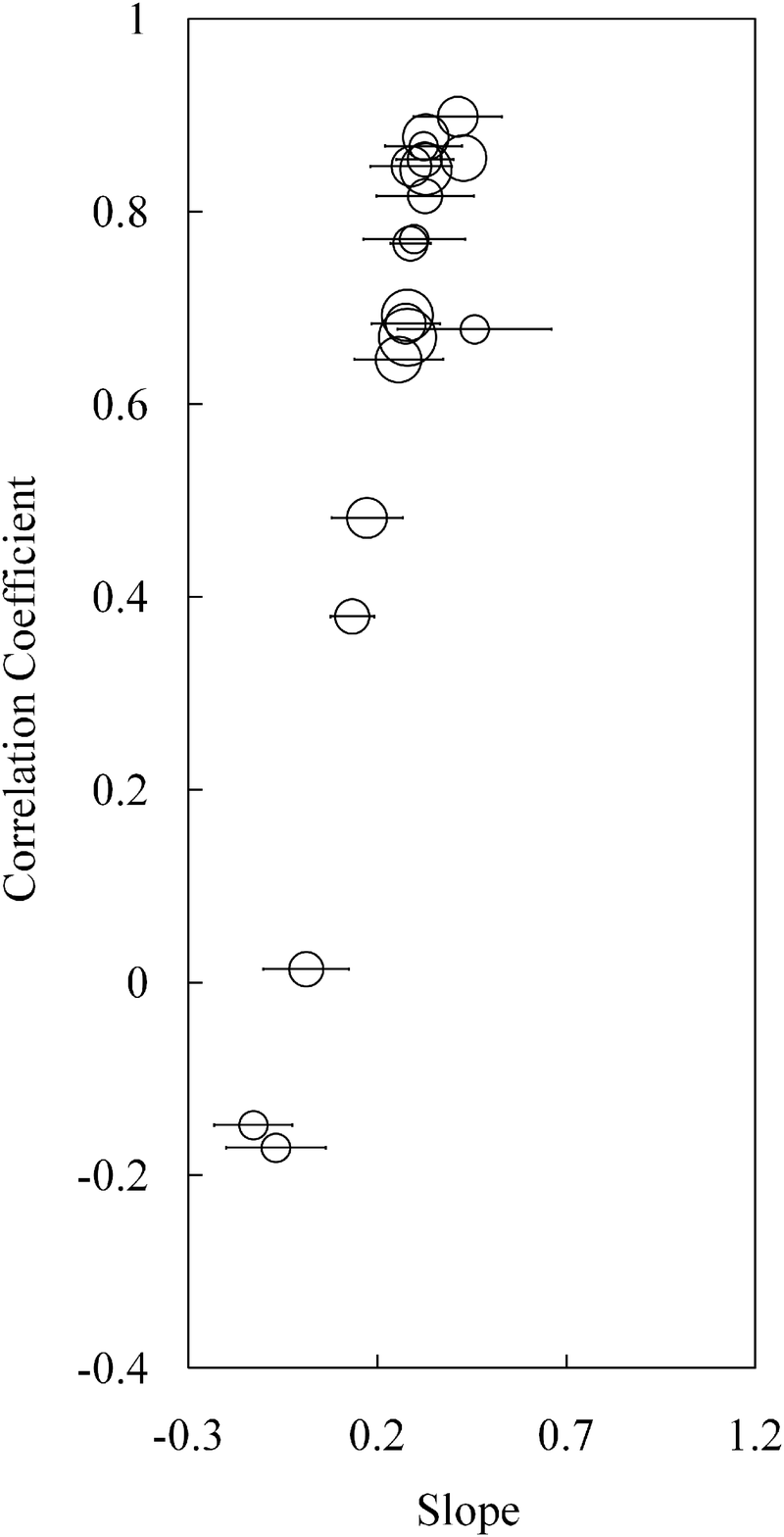}
        \caption{Cluster dissolution correction. The age bin in which the
                 effects of cluster number depletion are minimal is 10-20
                 Myr. In this figure we represent the correlation coefficient
                 as a function of the power index for different cluster 
                 subsamples in that age range (WEBDA data). Point symbol 
                 sizes are proportional to the size of the age-range used. 
                 Clusters located within 1 kpc from the Sun (left), 2 kpc 
                 (middle), and the full sample (right) are shown. If only 
                 the best fits are selected (correlation coefficient 
                 $>$ 0.71), the values of the slope cluster are $\sim$0.44 
                 (left), $\sim$0.40 (middle), and $\sim$0.33 (right). We will 
                 consider the cluster subsample located within 2 kpc from the 
                 Sun as the best sample, including both the cluster dissolution 
                 and incompleteness corrections. The value of the slope for 
                 this subsample is used to implement the corrections (see the 
                 text for details).
                 \label{correction}
                }
     \end{figure}

    Unfortunately, in both the analysis completed above and the one in Efremov
    \& Elmegreen (1998), the important effect of dissolving clusters has not 
    been taken into consideration. The number of open clusters decreases over 
    time because they merge, become unbound and dissolve after rapid mass 
    ejection, or they evaporate as a result of two-body relaxation or tidal 
    shocks with molecular clouds in the gravitational field of the Galaxy (see 
    de Grijs \& Parmentier 2007 for a recent review). The first two processes 
    (merging and infant mortality) are only relevant during the early evolution 
    of open clusters (clusters younger than 20-30 Myr), nearly 2/3 of clusters 
    disappear during this initial stage (e.g. de la Fuente Marcos \& de la Fuente 
    Marcos 2008). Beyond that age, the third process becomes dominant and over 
    a time-scale of 100 Myr, the surviving fraction of clusters (after the infant 
    mortality phase) is depleted by about 25\% (see de la Fuente Marcos \& de la 
    Fuente Marcos 2008 for a more detailed discussion). The role of cluster 
    dissolution and its impact on the correlation obtained above is obviously not 
    negligible; as clusters are depleted, the average intercluster separation 
    increases. Fortunately, the effect of cluster dissolution is rather small 
    in the age range 1-20 although incompleteness due to embedded 
    clusters is relevant in the age range 1-10 Myr. This younger age group
    is strongly affected by detectability issues as embedded proto-open 
    clusters may not be observable at optical wavelengths if the amount of internal 
    extinction due to gas and dust is too high. The situation described in star 
    complexes is similar to the one found in individual young open clusters where 
    pre-main sequence stars (many of them still embedded in their primordial gas 
    cocoons) are found together with fully functional main sequence stars. As a 
    correction to the result in Equation \ref{rel} is mandatory, it should be based 
    on the analysis of the open cluster population in the age range 10-20 Myr: the 
    one for which the effects of cluster disruption and incompleteness are less 
    relevant. In principle, such a correction has to preserve the power-law nature 
    of the correlation, being independent on the age binning considered and providing 
    a statistically relevant result. In order to implement this correction, we have 
    considered all the possible binning range combinations (5 to 10 Myr) within that 
    age interval and calculated the corresponding slope for each one of them. The 
    binning choice naturally reflects the level of uncertainty in young clusters age 
    determination. The result is displayed in Figure \ref{correction}, right panel. 
    The sizes of the result points are proportional to the age bin used.
    If we recalculate the average parameters of the $\Delta t-S$ relation using the 
    values with the highest correlation coefficient ($r > 0.71$, so at least 50\%
    of the variance in the average age difference can be explained by the regression
    equation), we obtain: 
    \begin{equation}
       \log \Delta t ({\rm Myr}) = -0.55\pm0.12 \ + 
                                   0.33\pm0.05 \ \log S({\rm pc}), \ \ \
               {\rm 10-20 \ Myr}\,,
    \end{equation}
    or
    \begin{equation}
       \Delta t ({\rm Myr}) \sim 0.28 \ S({\rm pc})^{0.33}\,.
        \label{relcor}
    \end{equation} 
    In other words, the overall effect of not accounting for the dissolving of 
    clusters is to make the apparent characteristic star formation time-scale 
    longer; i.e. star formation appears to be slower. This effect also affects
    the results in Efremov \& Elmegreen (1998). The actual power index for the
    LMC after correction may well be in the range 0.4-0.6. If all the clusters 
    in the age bin 10-20 Myr are considered and no averaging is attempted we 
    obtain:
    \begin{equation}
       \Delta t ({\rm Myr}) \sim 0.6 \ S({\rm pc})^{0.3}\,,
        \label{relcorother}
    \end{equation} 
    which is consistent with the previous determination but a factor 2 slower.
    The reasoning behind the averaging process is as follows. The peak in the
    open cluster age distribution is found at 15 Myr (de la Fuente Marcos \& 
    de la Fuente Marcos 2008). The embedded cluster phase lasts $\approx$ 10 
    Myr. After 20 Myr a sharp decline in cluster numbers is observed. On the
    other hand, the actual time-scales depend on the environmental conditions
    (location within the Galaxy). The averaging is an attempt to compensate
    for missing clusters because they have not yet been detected or because they
    have already been destroyed, the two ends of the age interval. It also 
    accounts for the large uncertainties in young cluster age determination,
    likely 50\% or more. It may however be argued that the 10-20 Myr data interval 
    results are not statistically significant (low number of open cluster pairs, 
    424 pairs or less for each calculated value of the slope in Figure \ref{correction}). 
    If we attempt a random trial experiment analogous to the one described
    above, we obtain similar values for the probability of this result being
    obtained by chance, $<$0.0005. Therefore, we will consider our result as
    statistically robust.

    The dissolution correction implemented above increases the value of the slope
    by nearly 50\% but as pointed out previously there is an additional effect that 
    may contribute to yield a smaller value for the slope and, therefore, a longer 
    characteristic star formation time-scale. Our original sample is contrast-induced, 
    magnitude-limited and some incompleteness is expected. A simple and robust approach 
    to minimize the effects of incompleteness consists in considering volume-limited 
    samples. In Figure \ref{correction} we display the equivalent findings (see above) 
    for a subsample of clusters located within 1 kpc (left panel) and 2 kpc (middle 
    panel) from the Sun. The slope increases ($\sim$20\%) with respect to the previous 
    correction. For young clusters, samples may be considered 90\% complete up to 2.5 kpc 
    from the Sun (see de la Fuente Marcos \& de la Fuente Marcos 2008, Figure 1). The number of
    cluster pairs in the entire sample for the age range 10-20 Myr is 424 and the 
    number for the 1 and 2 kpc samples are 140 and 336, respectively. Therefore and in
    order to preserve some statistical significance, we will consider the subsample found
    within 2 kpc from the Sun. Recalculating the average parameters of the $\Delta t-S$ 
    relation using again the values with the highest correlation coefficient ($r > 0.71$), 
    we now obtain: 
    \begin{equation}
       \log \Delta t ({\rm Myr}) = -0.7\pm0.2 \ + 
                                    0.40\pm0.08 \ \log S({\rm pc}), \ \ \
               {\rm 10-20 \ Myr}\,,
    \end{equation}
    or
    \begin{equation}
       \Delta t ({\rm Myr}) \sim 0.2 \ S({\rm pc})^{0.4}\,.
        \label{relcorfinal}
    \end{equation} 
    The rationale for adopting as our best subsample all clusters in the age interval 
    10-20 Myr and located within 2.0 kpc from the Sun is that in a volume-limited sample 
    and at these young ages, the vast majority of the open clusters present have been
    detected even in the presence of early gas expulsion. The 1 kpc subsample produces
    a slightly higher value for the slope, 0.44$\pm$0.13, but its error interval overlaps with that
    of the 2 kpc sample so we choose the 2 kpc value for being statistically more significant.
    The value found is a plausible lower limit for the correction factor. The apparently
    strong correlation observed in Figure \ref{correction} is a natural consequence of the
    fact that data with low correlation results in models with slope close to zero. On the
    other hand, if the slope is significantly different than zero, then the regression model
    can actually be of some use. The clusters of points around the {\it optimal} values 
    contribute to enhance this apparent correlation.

    Equation \ref{relcorfinal} strongly supports the idea of rapid star formation:
    for a 40 pc cloud, the time-scale for star formation is just 1 Myr. This
    result is consistent with works by Elmegreen (2000, 2007), Ballesteros-Paredes \& Hartmann 
    (2007) and Tamburro et al. (2009). The formation times in dense,
    bound clusters suggest that the main star formation activity is over in
    $\sim$3 Myr.  

  \section{The closest pairs}    
\begin{figure}
        \epsscale{0.85}
        \plotone{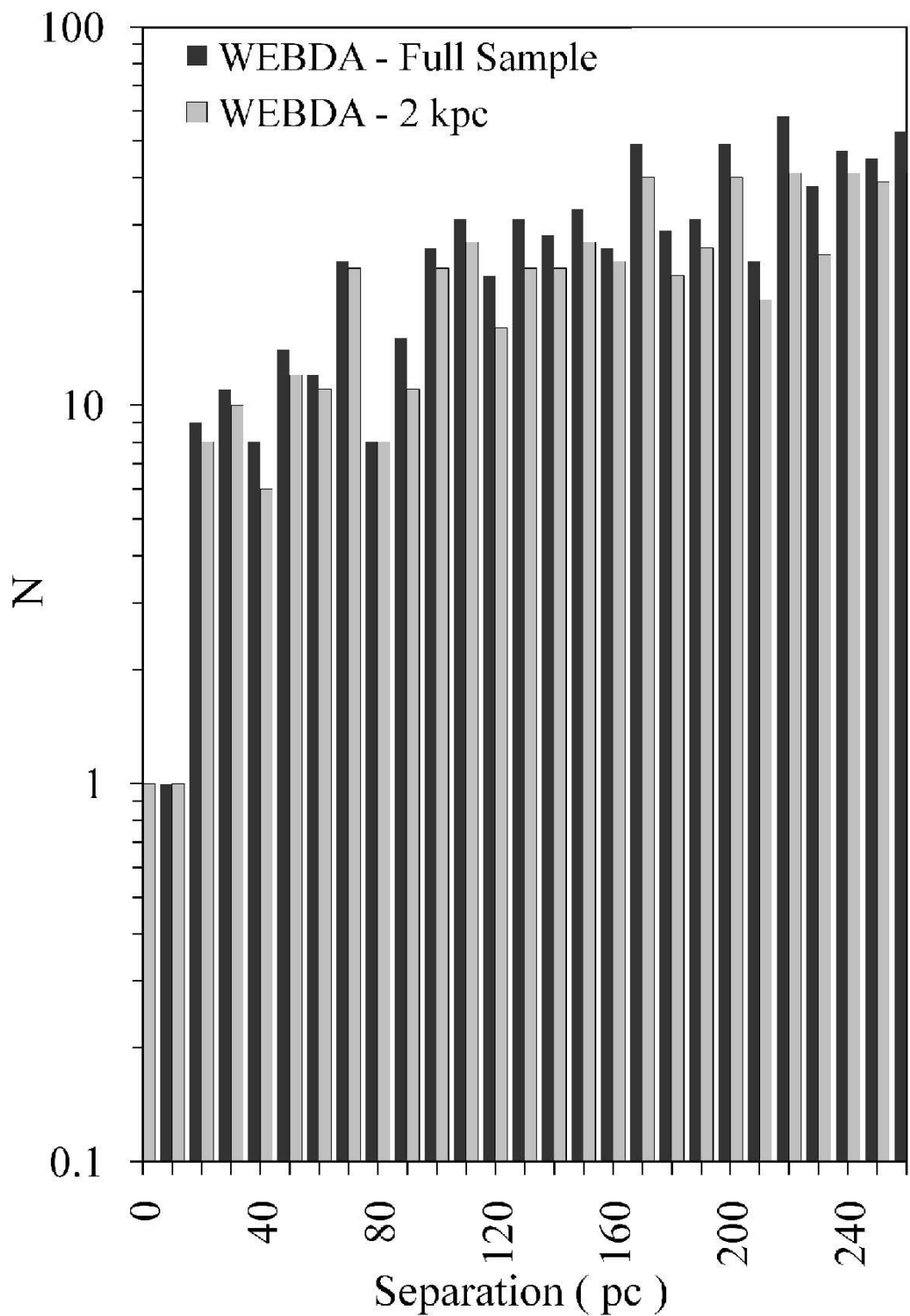}
        \caption{Intercluster distance histogram for young (age $<$ 45 
                 Myr) cluster pairs in WEBDA closer than the NGC 869/NGC 
                 884 separation (267 pc, WEBDA data). Two samples are 
                 displayed, the full one and the 2 kpc sample. The two 
                 histograms appear to be rather similar. The distance bin 
                 is 10 pc. Very few young pairs have separation $<$ 20 pc
                 (pairs \#1 and \#2). 
                 \label{distanceh}
                }
     \end{figure}

\begin{figure}
        \epsscale{0.85}
        \plotone{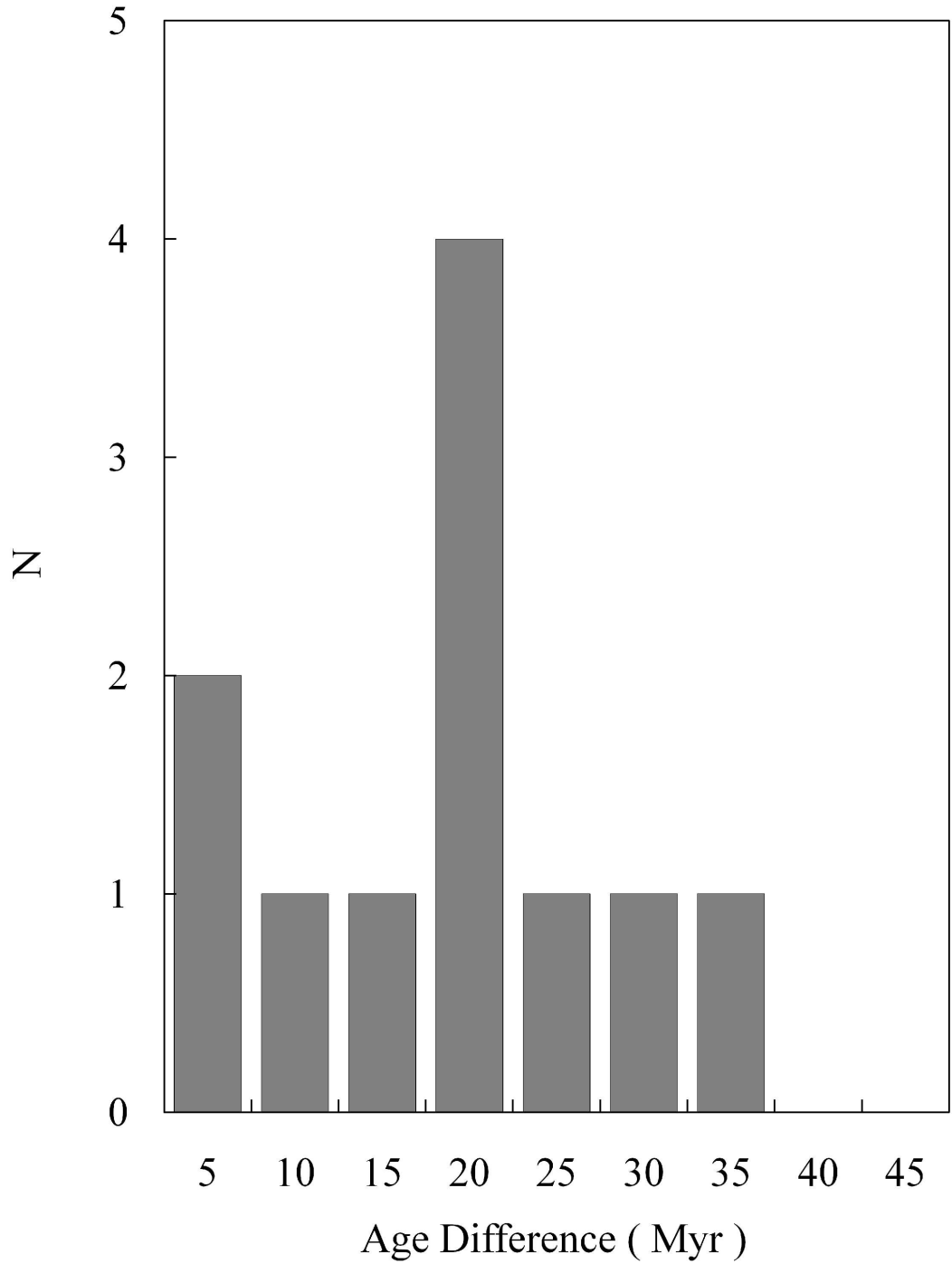}
        \caption{Age difference histogram for young (age $<$ 45 Myr) cluster
                 pairs in WEBDA closer than 30 pc (Table \ref{binaries}).
                 \label{ageh}
                }
     \end{figure}

    The existence of double/binary star clusters in the Clouds of Magellan
    was first proposed by Bhatia \& Hatzidimitriou (1988, LMC) and 
    Hatzidimitriou \& Bhatia (1990, SMC). Over the last few years, this result
    has been widely accepted (see Dieball et al. 2002 for a recent study). On 
    the other hand, attempts to identify binary open cluster candidates in the
    Milky Way have given a number of probable pairs (see Subramaniam et al.
    1995 for a catalog). To date, the only widely accepted double or binary
    open cluster in the Galaxy is the $h + \chi$ Persei pair (NGC 869/NGC 884).
    The actual physical (not projected) separation between the pair members 
    is, however, 267 pc (using WEBDA data). If they constitute a 
    bound system, its binding energy must be very low and it is highly unlikely
    to persist in this hypothetical binary state for a long period of time.
    The analysis carried out in the previous section has produced a number of
    open cluster pairs with separations significantly below the value found
    for the Double Cluster. In Table \ref{binaries} we compile a list of
    probable binary open clusters. The main criterion used in this compilation
    is purely dynamical: the pair separation must be less than three times
    the average value of the tidal radius for clusters in the Milky Way disk
    (10 pc, Binney \& Tremaine 2008). Following Innanen et al. (1972), for two
    clusters separated by a distance larger than three times the outer radius
    of each cluster, the amount of mutual disruption is rather negligible. This 
    simple criterion produces only 34 pairs (6.6\% of clusters in our full sample).
    Out of them, we select those with an age difference $<$ 45 Myr as they are
    the most likely candidates to be actual binary open clusters; they are very
    close and almost coeval. The cluster pair separation histogram for this
    young sample is displayed in Figure \ref{distanceh}. Here we only focus on
    cluster pairs closer than the Double Cluster separation. Only a few pairs are
    closer than 40-50 pc. Most of these pairs could be undergoing some type of 
    tidal interaction. The age difference distribution for the closest, likely
    interacting, subsample is shown in Figure \ref{ageh}. In an effort to rate the 
    strength of this interaction, we propose the following classification scheme 
    (\ref{bincri}) based only in the values of separation ($S$) and tidal radii 
    ($R_T$):
    \begin{equation}
       {\rm Cluster \ Pairs} \left\{ \begin{array}{l}
                                     {\rm Detached}, \ R_{T1} + R_{T2} < S \\
                                     \begin{array}{c}  
                                      {\rm Interacting} \\
                                      R_{T1} + R_{T2} > S
                                     \end{array}
                                      \left\{
                                       \begin{array}{l}
                                          {\rm Weak}, R_{T1} \ {\rm AND}
                                                      \ R_{T2} < S \\
                                          {\rm Semi-Detached}, \ R_{T1} 
                                                      \ {\rm OR} \ R_{T2} < S \\
                                          {\rm In-Contact}, \ R_{T1} 
                                                      \ {\rm AND} \ R_{T2} > S
                                       \end{array}
                                     \right.
                                   \end{array}
                           \right.
       \label{bincri}
    \end{equation}  
    In order to apply this criterion, we use tidal radii from Piskunov et al.
    (2008). These radii have been determined in a self-consistent way from a
    fitting of King's profiles. For young clusters, the values calculated by 
    Piskunov et al. (2008) show just small discrepancies with respect to 
    previous determinations by Lamers et al. (2005). The classical King's Law 
    (King 1962, 1966) is able to fit the overall density of globular clusters 
    very well and also the density profiles of some elliptical galaxies 
    moderately well. It is based on an extensive study of the distribution of 
    stars in globular clusters. The empirical King profile is described by the 
    expression: 
    \begin{equation}
       \Sigma(r) = \Sigma_o 
                    \ \ \Biggl(\frac{1}{\sqrt{1 + (r/R_C)^2}} -
                                \frac{1}{\sqrt{1 + (R_T/R_C)^2}}\Biggr)^2 \,,
    \end{equation}
    where $\Sigma_o$ is the central surface density, $R_C$ is a scale factor 
    commonly called the core radius and $R_T$ is the value of $r$ (the distance
    from the assumed cluster center) at which $\Sigma$ reaches zero, the 
    limiting or tidal radius. Using a King density profile to model the surface
    density of young open clusters is a normal practice but it presents two 
    main problems: it assumes implicitly that the system under study is bound 
    and the tidal radius may be underestimated due to cluster ellipticity. 
    Interacting clusters are expected to be elliptical in shape (de Oliveira et
    al. 2000; Carvalho et al. 2008). We refer the reader to the original paper 
    by Piskunov et al. (2008) for the details on how these shortcomings have 
    been addressed. If the value of the tidal radius $R_{Ti}$ is unknown, the 
    pair type is left blank. 

\begin{figure}
        \epsscale{0.80}
        \plotone{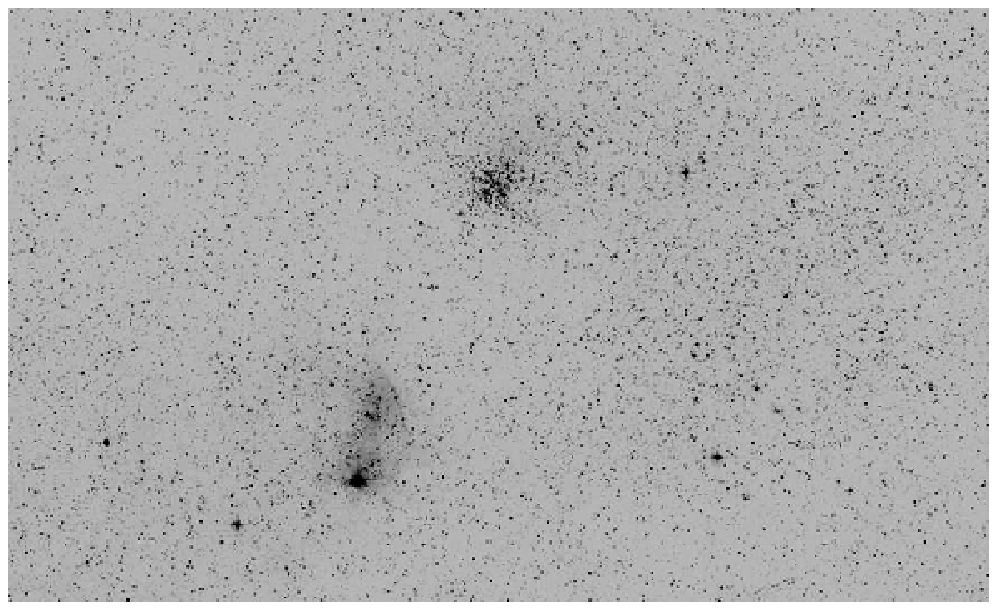}
        \plotone{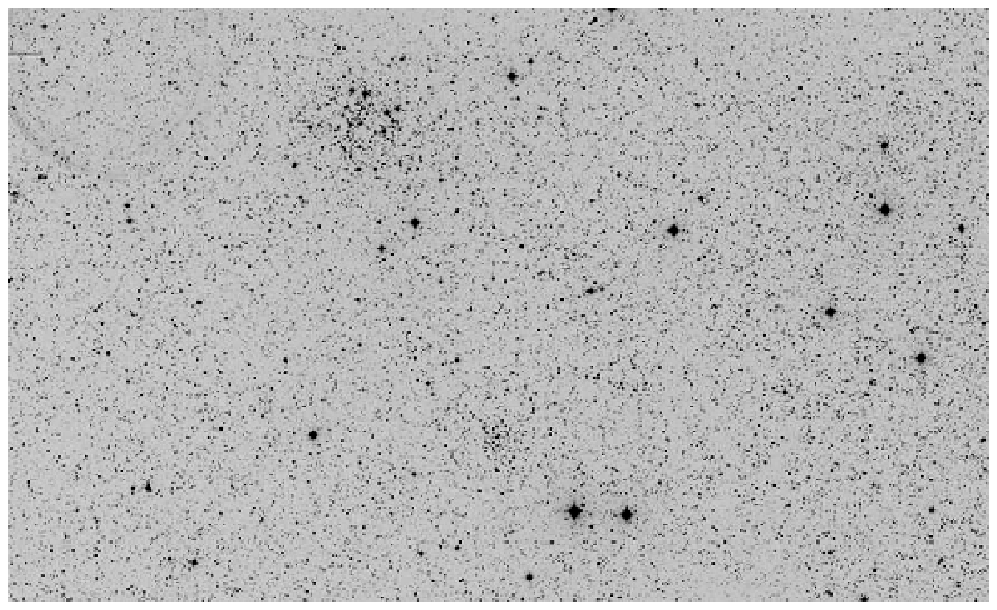}
        \caption{Some candidate binary open clusters from Table \ref{binaries}.
                 (Top) Pair \#5, NGC 3293/NGC 3324, a weakly interacting
                 pair near the $\eta$ Carinae nebula. NGC 3293, also known as 
                 the Gem cluster, is the object located on the upper-right
                 part of the frame and NGC 3324 appears at the bottom-left
                 (V-DSS1 frame, epoch=1987.05060651).
                 (Bottom) Pair \#8, NGC 659/NGC 663, another weakly 
                 interacting pair. NGC 663 is the object located on the 
                 upper-left part of the frame and NGC 659 appears at the 
                 bottom-center (E-DSS1 frame, epoch=1954.75111225188, Palomar
                 Observatory).
                 \label{examples}
                }
     \end{figure}

    No clusters in Table \ref{binaries} appear to be in-contact (the strongest 
    interaction level); therefore, it is either physically impossible or, more 
    likely, current observational techniques are unable to properly separate 
    young overlapping clusters. An alternative but a bit speculative possibility
    is that when close enough, the time-scale for subsequent merging or mutual 
    tidal destruction of clusters is very short (likely less than 16 Myr, the 
    average value of the age difference for pairs in Table \ref{binaries}). Only 
    one pair is candidate to be in the following level of significant 
    interaction, semi-detached: NGC 6618(M 17)/NGC 6613(M 18) in Sagittarius 
    by the Omega Nebula. Again, we must emphasize the strong observational 
    selection effect against identifying pairs of clusters in the semi-detached
    and in-contact categories. In contrast, the number of weakly interacting 
    clusters is relatively significant, 5 pairs. One of them may constitute a 
    candidate hierarchical triple system, with NGC 1976/NGC 1981 interacting 
    weakly and Collinder 70/NGC 1981 fully detached. NGC 1976 is the Orion 
    Nebula cluster or Trapezium cluster, Collinder 70 is the Orion's Belt 
    cluster. Also in Orion, we find the weakly interacting pair 
    ASCC 20/ASCC 16. NGC 3324/NGC 3293 is located in Carina (see Figure \ref{examples},
    top), NGC 663/NGC 659 is located in Cassiopeia (see Figure \ref{examples}, bottom), 
    and NGC 6871/Biurakan 1 is found in Cygnus. Besides the pair Collinder 70/NGC 1981, 
    the only other detached pair is ASCC 21/ASCC 18 also in Orion. The pair 
    ASCC 50/Collinder 197 is located in
    Vela, NGC 6250/Lynga 14 is in the border between the constellations Ara and
    Scorpius, and Trumpler 24/NGC 6242 is found in Scorpius. Not surprisingly, all 
    the proposed candidate binary open clusters are in the vicinity of well known OB 
    stellar associations and/or HII region complexes. The Orion Nebula neighborhood
    appears to be the region with the highest open cluster density and it is
    also the closest. All the candidates identified here are young and, 
    therefore, they are expected to retain an almost primordial kinematics 
    with similar radial velocity and proper motions within the pair unless we
    are witnessing an ejection after a close fly-by (as may be the case in
    pair \#8). Two pairs (\#5 and \#8) are displayed in Figure
    \ref{examples}. It is rather difficult to find public data in which any
    of the pairs in Table \ref{binaries} are portrayed together. In one of the 
    displayed cases, \#5, we clearly observe that the oldest cluster
    in the pair appears to be significantly more expanded. This is consistent 
    with findings by Baumgardt \& Kroupa (2007): after gas expulsion, surviving
    clusters have typically expanded by a factor of 3 or 4 due to gas removal. 
    This behavior was already predicted on purely theoretical grounds by Hills
    (1980).

  \section{Discussion}
    Star complexes appear to be the largest coherent star-forming units in the
    Milky Way disk and elsewhere. With characteristic diameters of about 600 pc,
    if we apply Equation (\ref{rel}), the intrinsic star formation time-scale for
    these structures is $\sim$30 Myr. Star clusters form in a shorter time-scale in 
    the LMC, this is correct even if we take into account the effects of cluster 
    dissolution and incompleteness. That may explain why, in general, young 
    clusters are more massive in the LMC. For size-scales $>$ 350 pc, the clustered 
    star formation time-scales in the Milky Way and the LMC are nearly the same: for 
    star complexes both are $\sim$30 Myr. If we accept the 
    results based only on the 10-20 Myr age range as statistically significant 
    and the cluster dissolution correction is included, the characteristic star
    formation time-scale in the quiescent Milky Way disk is dramatically 
    reduced: $\sim$ 1 Myr. This value supports the idea of rapid star 
    formation and is consistent with recent results compiled by 
    Ballesteros-Paredes \& Hartmann (2007): the median age of stars in nearby 
    molecular clouds is $\sim$1-2 Myr and the median age of star-forming 
    molecular clouds is also 1-2 Myr. On the other hand, the existence of a 
    relatively large number of almost equal-age pairs implies that star 
    formation is synchronized in neighboring regions, which means that there is
    only a short time interval available for the complete formation of a 
    cluster and its neighbor (Elmegreen 2000). Tamburro et al. (2009) have 
    recently found a short time-scale for the most intense phase of star formation
    in spiral arms, 1-4 Myr. A considerable fraction of giant molecular clouds
    exist only for a few Myr before forming stars. 
    
    Although statistically robust, our results are affected by the intrinsic
    errors associated to the determination of open cluster parameters. The 
    current status of the accuracy of open cluster data in the Milky Way has 
    been reviewed by Paunzen \& Netopil (2006). In this work, it is pointed 
    out that distances are rather well known because for about 80\% of 395 of 
    the best studied objects, the absolute error is $<$ 20\%. The situation 
    is just the opposite for ages, with only 11\% of the investigated open 
    clusters (best sample again) having errors $<$ 20\% and 30\% with absolute 
    errors $>$ 50\%. Besides, errors are age dependent and the ones associated 
    to young clusters are sometimes significantly larger (in average) than 
    those of older objects (see their list of 72 suggested standard open 
    clusters): 20-30\% in the best possible cases. In a large sample, these 
    errors are very likely to be non-homogeneous as different methods have been
    used by different authors to calculate the ages. In order to check for 
    consistency, we have also used the {\it New Catalogue of Optically Visible 
    Open Clusters and Candidates} (NCOVOCC)\footnote{http://www.astro.iag.usp.br/$\sim$wilton/}.
    The first version of this catalogue was presented in Dias et al. (2002). 
    The February 2009 version (v2.10, Dias 2009) includes 1787 objects, with distances 
    and ages for 982 clusters. If we repeat the calculations using the values provided by 
    NCOVOCC for the full sample we obtain consistent values:
    \begin{equation}
       \log \Delta t ({\rm Myr}) = 1.12\pm0.04 \ + 
                                   0.146\pm0.016 \ \log S({\rm pc}), \ \ \
               {\rm 1-100 \ Myr, r = 0.94}\,,
    \end{equation}
    \begin{equation}
       \log \Delta t ({\rm Myr}) = 1.17\pm0.04 \ + 
                                   0.125\pm0.015 \ \log S({\rm pc}), \ \ \
               {\rm 10-100 \ Myr, r = 0.93}\,,
    \end{equation}
    \begin{equation}
       \log \Delta t ({\rm Myr}) = 1.80\pm0.07 \ + 
                                   0.22\pm0.03 \ \log S({\rm pc}), \ \ \
               {\rm 1-1000 \ Myr, r = 0.93}\,.
    \end{equation}
    The corrected value for the slope is now 0.42$\pm$0.07 which is also consistent 
    with the value obtained using data from WEBDA. Unfortunately, the situation is 
    significantly worst for the LMC where, in addition to the problems with age
    determinations, the issue of computing the actual physical distances to 
    individual clusters is complicated by the nature of the host galaxy. In
    summary and if an optimistic perspective is adopted, all the numerical 
    results obtained in this research may well be affected by errors of 20\%. 
    On the other hand and strictly speaking, our results are only applicable to
    current clustered star formation at the Solar Circle.

    Regarding the presence of binary open clusters in the Milky Way disk, we
    conclude that there is a small population of candidate binary (and perhaps
    higher multiplicity) open clusters. Our list of candidates in Table
    \ref{binaries} has only 1 pair in common with the classical study by
    Subramaniam et al. (1995): Collinder 70/NGC 1981 in Orion. This is to be
    expected as they used projected distances instead of physical separations.
    If open clusters are born in complexes with characteristic sizes of
    600 pc, some clusters formed along the shock fronts induced by supernova 
    explosions and all of them observed in projection, the likelihood of 
    misidentifying large numbers of double but not physically close young clusters 
    as actual binaries is very high. As clear examples of this situation let us 
    consider the $h + \chi$ Persei pair discussed before or the double cluster 
    NGC 1912/NGC 1907 identified by de Oliveira et al. (2002) as two clusters 
    experiencing a fly-by (but the actual separation is 490 pc). 
    One of the most intriguing properties of the sample in Table \ref{binaries}
    is the relatively large number (62\%) of pairs with large differences in the
    values of their tidal radii. According to recent simulations carried out
    by Portegies Zwart \& Rusli (2007), cluster binary orbits tend to expand
    due to mass loss but the initially less massive cluster expands more
    quickly than the binary separation increases. Therefore and contrary to
    the stellar binary case, the less massive cluster tends to start mass
    transfer to the most massive cluster. Our results suggest that this process
    may already be happening within the pair NGC 6618/NGC 6613 as they
    exhibit the largest ratio of tidal radii ($>$ 5) and the tidal radius
    of NGC 6618 is almost twice the pair separation. This system is probably
    bound to merge within the next few Myr. Besides, gas expulsion has also an
    effect on star cluster evolution; as pointed out before, Baumgardt \& Kroupa
    (2007) have found that after gas removal the surviving cluster expands 
    significantly although Hills (1980) was first in predicting this evolutionary 
    behavior as well as early cluster disruption. In two cases the age difference 
    is so small that the pairs are likely to be primordial, i.e. they were born 
    {\it twins}: ASCC 18/ASCC 21 and NGC 3293/NGC 3324. In the other nine cases 
    and taking into consideration the errors in age determination, the pair could 
    be the result of capture in a high cluster density environment (the open 
    cluster family, see de la Fuente Marcos \& de Fuente Marcos 2008, 2009). In 
    any case, the observed separations are unlikely to be primordial and they may 
    have increased to produce eccentric orbits as described in Portegies Zwart \& 
    Rusli (2007). Regarding the observational selection effect against identifying 
    pairs of clusters in the two highest levels of interaction, the recent results 
    on Bochum 1 may shed some light on this issue. The area around Bochum 1 and 
    FSR911 may be considered as a candidate example for in-contact pairs (see Bica 
    et al. 2008 for details).

  \section{Conclusions}
    Our main conclusions can be summarized as follows:

    (0) In the Milky Way, open clusters with large spatial separation also 
        tend to have greater age separations. This result is independent on 
        the age group. The correlation found is similar to the one identified 
        for the LMC. 

    (1) Star formation in the Milky Way disk proceeds hierarchically.

    (2) The average age difference between pairs of open clusters in
        the Milky Way disk increases with their separation as the 
        $\sim$0.16 power. This result is obtained for the 1-100 Myr 
        age interval and the 0-600 pc distance range. If the effects
        of open cluster dissolution and incompleteness are corrected 
        by using volume-limited samples of clusters in the age interval 
        10-20 Myr, the value of the power index increases to about 0.4.
 
    (3) Open cluster dissolution, if not properly accounted for, induces 
        a significant overestimate in the characteristic star formation 
        time-scale. 

    (4) Quiescent star formation in the Milky Way disk appears to be
        very rapid with a characteristic time-scale of about 1 Myr.

    (5) The vast majority of open clusters appear to form within some
        type of complex.

    (6) The Milky Way disk appears to host a small but non-negligible
        population of candidate binary open clusters.

    (7) Components in a binary (or higher multiplicity) open cluster
        are likely formed together. This is a confirmation of the
        result first obtained by Fujimoto \& Kumai (1997) and later
        supported by the analysis in Dieball et al. (2002). 

    (8) Identification of in-contact pairs, both $R_{T1}$, $R_{T2}$ $<$ 
        $S$, is an extreme observational challenge. This may explain the 
        absence of candidates of this type in our data set. An alternative
        but less likely interpretation may be that fully overlapping open
        clusters are extremely unstable with merging or tidal destruction 
        time-scales $<$ 16 Myr, the average age difference for pairs in Table 
        \ref{binaries}. 

    (9) Binary cluster candidates in the Galactic disk appear to show a
        tendency to have components of different size (5 out 8 studied pairs
        have size difference $>$50\%). This is likely the result of dynamical 
        interaction but it could also be caused by an observational selection 
        effect against identifying large, not concentrated interacting 
        clusters. 

  \acknowledgments
     We thank Charles Bonatto and Eduardo Bica for their comments on Bochum 1.
     In preparation of this paper, we made use of the NASA Astrophysics Data 
     System and the ASTRO-PH e-print server. This research has made use of the 
     WEBDA database operated at the Institute of Astronomy of the University of
     Vienna, Austria.

   \clearpage

         \begin{table}
          \fontsize{8} {10pt}\selectfont
          \tabcolsep 0.10truecm
          \caption{List of candidate binary open clusters}
          \begin{tabular}{rccrrrrrrc}
           \hline
           \multicolumn{1}{c}{Pair \#}        &
           \multicolumn{1}{c}{Cluster 1}      &
           \multicolumn{1}{c}{Cluster 2}      &
           \multicolumn{1}{c}{$\tau_1$}       &
           \multicolumn{1}{c}{$\tau_2$}       &
           \multicolumn{1}{c}{$\Delta t$}     &
           \multicolumn{1}{c}{$R_{T1}$}       &
           \multicolumn{1}{c}{$R_{T2}$}       &
           \multicolumn{1}{c}{$S$}            &
           \multicolumn{1}{c}{Pair}            \\
           \multicolumn{1}{c}{}               &
           \multicolumn{1}{c}{}               &
           \multicolumn{1}{c}{}               &
           \multicolumn{1}{c}{(Myr)}          &
           \multicolumn{1}{c}{(Myr)}          &
           \multicolumn{1}{c}{(Myr)}          &
           \multicolumn{1}{c}{(pc)}           &
           \multicolumn{1}{c}{(pc)}           &
           \multicolumn{1}{c}{(pc)}           &
           \multicolumn{1}{c}{type}            \\
           \hline
             1  &  NGC 1981  &  NGC 1976  &  31.6  &  12.9  &  18.7  &   
                   3.9 & 6.6 & 7 & W \\ 
             2  &  ASCC 20   &  ASCC 16   &  22.4  &   8.5  &  13.9  &  
                   12.6 & 9.1 & 13 & W \\ 
             3  &  ASCC 50   & Collinder 197 &  30.2  &  13.4  & 16.8 & 
                   12.0 &  - & 20.5 & - \\
             4  &  NGC 6250  &  Lynga 14  &  26.0  &   5.2  &  20.8  & 
                   12.1 &  - & 20.7 & - \\ 
             5  &  NGC 3293  &  NGC 3324  &  10.3  &   5.7  &   4.7  & 
                   7.8 & 18.8 & 21.0 & W \\ 
             6  &  NGC 6613  &  NGC 6618  &  16.7  &   1.0  &  15.7  & 
                   7.3 & 38.2 & 21.9 & SD \\ 
             7  &  NGC 1981  &  Collinder 70 & 31.6 & 5.1 &  26.5  & 
                   3.9 & 20.3 & 24.5 & D \\ 
             8  &  NGC 659   &  NGC 663   &  35.3  &  16.2  &  19.1  & 
                   7.4 & 23.4 & 24.7 & W \\ 
             9  &  ASCC 18   &  ASCC 21   &  13.2  &  12.9  &   0.3  & 
                   11.5 & 13.7 & 25.4 & D \\ 
            10  &  NGC 6242  &  Trumpler 24 & 40.6  &   8.3  &  32.3  & 
                   9.8 & - & 25.5 & - \\ 
            11  &  Biurakan 1 &  NGC 6871  &  17.8  &   9.1  &  8.7  & 
                   13.5 & 14.2 & 27.4 & W \\ 
           \hline
          \end{tabular}
          \begin{list}{}{}
            {\footnotesize
              \item[] $\tau_i$: cluster age in Myr (WEBDA, $i$ = 1, 2). 
              \item[] $\Delta t = \tau_1 - \tau_2$: age difference in Myr. 
              \item[] $R_{Ti}$: cluster tidal radius in pc (Piskunov et al. 
                                2008, $i$ = 1, 2). 
              \item[] $S$: cluster pair spatial separation in pc. 
            }
          \end{list}
          \label{binaries}
         \end{table}

   \clearpage

%
%
%
%
%

   \clearpage

%
%
%
%
%

   \clearpage

%
%
%
%
%

   \clearpage

%
%
%
%
%

   \clearpage

%
%
%
%
%

   \clearpage

%
%
%
%
%

   \clearpage

%
%
%
%
%

\end{document}